\documentclass{nature}

\usepackage[utf8]{inputenc}
\usepackage{graphicx}
\usepackage{aa_sym}

\usepackage{amssymb}
\usepackage{amsmath}
\usepackage{mathrsfs}
\usepackage{color}

\title{Fast Inflow Directly Feeding Black Hole Accretion Disk in Quasars \protect\\
}

\author{Hongyan Zhou$^{1,2,3}$, Xiheng Shi$^{1,2}$, Weimin Yuan$^{4,5}$, Lei Hao$^{6}$, Xiangjun
Chen$^{2,3}$, Jian Ge$^{7}$, Tuo Ji$^{1,2}$, Peng Jiang$^{1,2}$, Ge
Li$^{2,3}$, Bifang Liu$^{4,5}$, Guilin Liu$^{2,3}$, Wenjuan Liu$^{8,9}$,
Honglin Lu$^{2,3}$, Xiang Pan$^{1,2}$, Juntai Shen$^{6}$, Xinwen
Shu$^{10}$, Luming Sun$^{2,3}$, Qiguo Tian$^{1,2}$, Huiyuan Wang$^{2,3}$,
Tinggui Wang$^{2,3}$, Shengmiao Wu$^{1,2}$, Chenwei Yang$^{1,2}$, Shaohua
Zhang$^{1,2}$ \&~Zhihao Zhong$^{2,3}$}

\begin{document}
\setlength{\unitlength}{1mm}
\sloppy

\maketitle

{\footnotesize
\begin{affiliations}
\item{Antarctic Astronomy Research Division, Key Laboratory for Polar Science of the State Oceanic Administration, Polar Research Institute of China, Shanghai, China}

\item{School of Astronomy and Space Sciences, University of Science and Technology of China, Hefei, China}

\item{Key Laboratory for Research in Galaxies and Cosmology of Chinese Academy of Sciences, Department of Astronomy, University of Science and Technology of China, Hefei, China}

\item{National Astronomical Observatories of China, Chinese Academy of Sciences, Beijing, China}

\item{School of Astronomy and Space Science, University of Chinese Academy of Sciences, Beijing, China}

\item{Shanghai Astronomical Observatory, Chinese Academy of Sciences, Shanghai, China}

\item{Department of Astronomy, University of Florida, Bryant Space Science Center, Gainesville, FL, USA}

\item{Yunnan Observatories, Chinese Academy of Sciences, Kunming, China}

\item{Key Laboratory for the Structure and Evolution of Celestial Objects, Chinese Academy of Sciences, Kunming, China}

\item{Department of Physics, Anhui Normal University, Wuhu, China}
\end{affiliations}

}

\begin{abstract}

Quasars are high-luminosity active galactic nuclei believed to be
powered by accretion of interstellar matter onto a super-massive
black hole (SMBH) therein. Most of the observed energy is released
in an accretion disk of inspiralling gas surrounding the SMBH. An
enormous amount of fueling material, up to several tens solar masses
per year, is expected to be transported inwards and consumed in the
end. However, basic questions remain unanswered as to whether and
how the accretion disks are supplied with external gas, since no
disk-feeding inflow has hitherto been observed clearly. Here we
report the discovery of highly redshifted broad absorption lines
arising from neutral hydrogen and helium atoms in a small sample of
quasars. Their absorption troughs show a broad range of Doppler
velocities from zero extending continuously inward up to as high as
$\sim 5,000$\,km\,s$^{-1}$, comparable to the free-fall speeds close
to the SMBH and constraining the fastest in-falling gas to be within
$10^4$ gravitational radii. We thus see through streams of cold gas
moving with a radially inward velocity component that spans an
immense gradient---a result of gravitational acceleration by the
central SMBH. Extensive photo-ionization modeling for the
archetypical object SDSS\,J103516.20+142200.6 indicates the
inflowing gas to be dense, thick and moderately ionized, with a
characteristic distance to the SMBH of $\sim 1,000$ gravitational
radii, possibly overlapping or close to the outer accretion disk.
Our results present the first compelling evidence for the
long-sought inflow directly feeding quasars' accretion disks with
external materials, likely originating from the dusty torus at a
parsec scale. Our approach provides a new tool to probe the bulk of
the so far elusive fueling inflows in quasars. Their studies may
help address some of the fundamental questions concerning accretion
physics, the onset and sustainment of quasar activity, and the SMBH
growth at centers of most galaxies.

\end{abstract}

\clearpage

Being the most luminous steady beacons known in the universe,
quasars have been a long-standing enigma in modern astrophysics
since their discovery\cite{Schmidt1963}. They are believed to be
powered by the gravitational potential energy of accreted matter
falling onto a SMBH with mass ranging from $M_{\bullet}\sim 10^{6}$
to $10^{10}~M_{\odot}$ at centers of galaxies. The energy release
takes place within an accretion disk of inspiralling gas in the
close vicinity of the SMBH extending out to a few $10^{2-3}$
gravitational radii ($R_{\rm g} \equiv GM_{\bullet}/c^2$). The observed
energetics require a substantial supply of mass accretion from a
fraction to several tens of solar masses per year\cite{Netzer2013}.
Despite the success of this paradigm since its first clear putting
forward half a century ago\cite{Lynden-Bell1969}, fundamental
questions remain unanswered: whether and how the accretion disks are
supplied with external gas? The answers are essential for
understanding further questions as to what sets off, maintains, and
terminates the quasar activity and how long the quasar phase lasts
for\cite{Krolik1999}. However, such an inflow directly supplying the
bulk of the material required to feed the accretion disks (termed
disk-feeding inflow hereafter), and subsequently the SMBHs, has
never hitherto been observed clearly, albeit some controversial
attempts in previous studies\cite{Hu2008,Sulentic2012}.

Atomic absorption lines imprinted on the observed spectra of quasars
are a robust probe of the dynamics and physical conditions of gas
along the line of sight (LOS) to the central radiation source. In
the optical--ultraviolet spectra of some $\sim 15\%$ quasars, broad
absorption lines (BALs) with velocity dispersions $\gtrsim
2,000~{\rm km~s}^{-1}$ by definition have long been observed, which are
mostly produced by resonant transitions of alkaline-like metals
including N\,V, C\,IV, Si\,IV, Al\,III and Mg\,II\cite{Crenshaw2003}. However,
their ubiquitous blueshifts (with respect to the quasar systemic
redshift) indicate outflows, rather than inflows, intrinsic to the
quasar central engine\cite{Lu2008}. Only recently BALs of metal ions
showing both blueshifted and redshifted components have been
observed in a handful of quasars\cite{Hall2013}. Nevertheless, it is
not yet clear to connect them to disk-feeding inflows, given the
large uncertainties in the quasar systemic redshifts (several times
$10^{2-3}$\,km\,s$^{-1}$, as estimated from high-ionization broad
emission lines) and a lacking of other evidence. In fact, metal BALs
are of little diagnostic value in probing cold and dense gas with
high column densities (subject to serious saturation and line
blending effects; see \textbf{Method} for details), which are
thought to be most probably the case of disk-feeding
inflows\cite{Shi2016}, if existing.

Motivated by the fundamental questions above, we have developed a
novel approach to diagnosing dense BAL gas of high column densities
and low/moderate ionization, based on extensive studies of our
own\cite{Zhang2015,Liu2015,Sun2017,Shi2017a,Zhang2018} and other
teams\cite{Hall2007,Leighly2011}. The approach makes use of BALs of
two new atomic series that became to know in quasar studies mostly
in recent years: the Balmer lines of neutral hydrogen
(H$_{n+2}~\equiv$
H$_{...,\zeta,...,\alpha}\lambda\lambda...,3890,...,6564$) and the
multiplets of neutral helium He\,I$^*_{n}$
(He\,I$^*_{...,3,2,1}\lambda\lambda...,3189,3889,10830$) arising from
transitions from the highly meta-stable $2~^3$S level. The Balmer
lines are mostly populated by collisional excitation in the neutral
zone of dense gas clouds/flows (with a density $n_{\rm H}\gtrsim
10^{6}$\,cm$^{-3}$), while the helium multiples populated via
recombination of He$^+$ ions. The relative abundances of H$_{n+2}$
and He\,I$^*_{n}$ are so low that some of their weak member lines
remain unsaturated even for Compton thick absorbers. Since these
line series are produced by transitions from the same low energy
level but with diverse oscillator strengths, we can decouple the
covering factor and column density in modeling the lines. The Balmer
lines depend on both the ionization and density, while the He lines
are sensitive almost solely to the ionization
parameter\cite{Ji2015}. We can therefore break the degeneracy of the
ionization and density by joint analysis of the two line series.
Lastly, the wide spaces in wavelength between the member lines
relieve the line-blending effect---a problem often encountered in
metal BALs.

In our ongoing program of systematic search for redshifted BALs of
both H$_{n+2}$ and He\,I$^{*}_{n}$, we discovered eight such objects
from $\sim 10^{5}$ quasars at redshifts $z<1.3$ in the Sloan Digital
Sky Survey (SDSS) spectral database of the latest fourteenth data
release (DR14)\cite{Paris2017}. We also obtained high
signal-to-noise optical and near-infrared spectra for SDSS
J103516.20+142200.6 (hereafter J1035+1422)---the archetype of the
sample---and several others via observations with the Paloma
200-inch telescope and the Keck~II 10-meter telescope, as well as
from the VLT/X-shooter spectral archive. The combined spectrum and
the BALs resulting from the analysis (see \textbf{Method}) for
J1035+1422 are presented in Figure\,\ref{Figure1_spectrum} and
\ref{Figure2_BALs}, while those for the remaining seven quasars in
Extended Data Fig. \ref{Fig3_sample} and \ref{Fig4_sample} in
\textbf{Method}. Although the H$_{n+2}$ BALs are superimposed on the
corresponding broad emission lines (BELs), we could disentangle one
from another by taking advantage of the Balmer decrements. We also
detected redshifted BALs of He\,I$^*_{1-3}$, after removing the
H$_{\zeta}$ BAL component. The redshifted BAL troughs of H$_{n+2}$
and He\,I$^{*}_{n}$ in the same objects show a largely common Doppler
velocity structure starting from $v\sim 0$ and extending
continuously up to a maximum velocity  that falls within $v_{\rm
max} \sim3,000 - 5,000$\,km\,s$^{-1}$ in five objects while $v_{\rm
max} \sim 1,000 - 1,500$\,km\,s$^{-1}$ in the remaining three (thus
classified as mini-BAL traditionally).

These detections are unprecedented. The lines are highly
Doppler-redshited and broadened by radially inward motion of the
absorbing gas with a large velocity gradient projected along the LOS
toward the central continuum source (nearly toward the SMBH). This
indicates that we are seeing through continuous streams of gases
that are transported inward to the central engine. The large maximum
BAL velocities measured are remarkable, being one order of magnitude
larger than those of starburst-driven outflows and winds of other
stellar processes\cite{Cicone2014}, suggesting the association with
the central engine. Since the radial velocity cannot exceed the
free-fall speed of the gravitational acceleration due to the central
SMBH, $v_{\rm max} < v_{\rm ff}(R) = \sqrt{2GM_{\bullet}/R}$ (from
infinity), we can constrain the distance of the innermost gas to the
SMBH as $R < 2GM_{\bullet}/v_{\rm max}^2 = 2R_{\rm g}(c/v_{\rm max})^2$.
For the majority of the sample we have $R \lesssim 10^4\,R_{\rm g}$,
well within the sphere of influence of the central black hole at the
order of $\sim 10^6\,R_{\rm g}$ scale. Specifically for J1035+1422,
$R \lesssim 7,200\,R_{\rm g}$. The distance limit would become even
smaller if the kinematics of the gas deviates from sole free-fall
motion (e.g.\ subject to a centrifugal force due to sub-Keplerian
motion and/or strong radiation pressure from the central source), or
there exists even faster infalling part of the inflow not
intersecting the LOS. Inevitably, to gain such high radial
velocities, the gases must be experiencing gravitational
acceleration by the central SMBH. The large velocity gradient
suggests that the LOS intersects a large number of (most likely
inspiralling) streams at various radii across a wide radial
distance, starting from somewhere at larger radii and ending much
closer inside.

In the rest of this paper, we performed further detailed analysis
for J1035+1422 ($z=1.2528\pm 0.0001$) as demonstration, which has
the highest data quality among the sample. The quasar has a mass
accretion rate $\dot{M}_{\odot}\approx 13\,M_{\odot}\,{\rm yr}^{-1}$
derived from its luminosity using the standard bolometric correction
($L_{\rm bol}\simeq 8.1\,L_{5100}$)\cite{Kaspi2005} assuming an
accretion efficiency $\eta = 0.1$, and a size of the broad emission
line region (BELR) $R_{\rm BELR}\approx 0.33$\,pc derived from the
$R_{\rm BELR}$--luminosity relation\cite{Kaspi2005}. We estimated
the black hole mass $M_{\bullet}\approx 2\times10^{10}\,M_{\odot}$
from the parameters of the reconstructed Balmer BEL and the size of
the BELR, with an uncertainty of $\sim$0.5\,dex\cite{GreeneHo2005}.
This leads to a gravitational radius $R_{\rm g} \approx 9.8\times
10^{-4}$\,pc and an Eddington ratio $\eta\equiv
L_{\rm bol}/L_{\rm Edd}\approx 0.03$.

Redshifted BAL troughs of metal ions (including C\,IV, Al\,III,
Mg\,II, and resonant and excited Fe\,II) were also found in the UV
part of the spectrum, showing similar velocity structures to those
of H$_{n+2}$ and He\,I$^{*}_{n}$ (Figure\,\ref{Figure2_BALs}; see also
Extended Data Fig.\,\ref{Fig7_CIV} and \ref{Fig8_UVFeII} in
\textbf{Method}). These metal BALs show an additional blueshift
component that is ubiquitously seen in normal BAL quasars. A similar
blueshift component is also seen in the He\,I$^*_{1}$ BAL. This
indicates the co-existence of an outflow with the inflow.
Surprisingly, the redshift component of the C\,IV BAL is much weaker
in strength than those of the H$\alpha$ and Mg\,II BAL. This is in
contrast to what is commonly observed in typical quasar
(blueshifted) BALs\footnote{$^{\dag}$For any gas directly exposed to
ionization continuum, $N_{C^{3+}}$ reaches the maximum at a much
shorter distance along the line of sight from the illumination
surface than $N_{Al^{2+}}$, $N_{Mg^{+}}$, $N_{He^{+}}$ and
$N_{HI^{*}(n=2)}$. This makes the C\,IV BAL easily saturated even
when the BALs of Al\,III, Mg\,II, He\,I$^*$ and H$\alpha$ remain
undetectable, as commonly shown in observations.}$^{\dag}$, and can
well be explained if the central ionizing radiation is filtered
prior to illuminating the inflow gas, naturally by the blueshifted
BAL outflow.

We measured the covering factor $C_{\rm f}(v)$ and the column densities
of $n=2$ neutral hydrogen $N_{{\rm H\,I}(n=2)}(v)$ and further of $2~^3$S
neutral Helium $N_{{\rm He\,I}^*(2~^3{\rm S})}(v)$ ($\propto$ $N_{{\rm H\,I}(n=2)}(v)$) as
a function of velocity. To constrain the physical properties of the
inflow, we performed extensive photo-ionization simulations and
compared the results with the redshifted BAL spectra observed (see
\textbf{Method} for details). The best-fit model indicates that the
inflow gas is relatively dense, thick and  moderately ionized (a
density $\log\,n_{\rm H} = 7.20^{+0.40}_{-0.45}~{\rm cm}^{-3}$, a total H
column density $\log\,N_{\rm H} = 23.46^{+0.33}_{-0.24}~{\rm cm}^{-2}$ and
an ionization parameter $\log\,U = -0.24^{+0.06}_{-0.64}$ at the
90\%~confidence level). We found the most probable characteristic
distance of the inflow to the SMBH, as represented by the optical
depth-weighted distance, to be $R_{\rm IF} =
1.01^{+2.16}_{-0.50}$\,pc (Figure\,\ref{Figure3_probability}, see
also Extended Data Fig.\,\ref{Fig6_prob} in \textbf{Method}). The
distance corresponds to $R_{\rm IF} \simeq 525-3,236\,R_{\rm g}$,
complying with the above upper limit constrained by the simple
physics of gravity. It is worth noting that, albeit large
uncertainties, the inner inflow has an inferred radial distance
comparable to the outer radius of the accretion disk in J1035+1422,
which we estimated to be from a few hundred to about one thousand
gravitational radii. The actual innermost end of the downstream may
be at even smaller radii if it deviates from---and does not
intersect---the line of sight. We calculated the force exerted on
the gas by the central radiation to be only $1-2\%$ of the black
hole's gravity. Hence the dynamics of the inflow gas is dominated by
the SMBH.

These properties leave little room for the other destinies of the
inflow but the one as the following. The bulk of the inflow
downstream would mostly be gravitationally pulled onto a general
plane defined by the overall angular momentum, as depicted in
Figure\,\ref{Figure4_cartoon}. Ultimately, the materials transported
therein by the inflow are likely somehow dumped into the outer
accretion disk and/or the marginally unstable region extending
further out. Assuming an axial symmetric geometry on the general
plane for the global inflow, we estimated a total mass inflowing
rate from the above derived total column density, $\dot{M}_{\rm
IF}\sim 30\,M_{\odot}$\,yr$^{-1}$ (see \textbf{Method}). This is
sufficient to power the quasar radiation luminosity observed ($\sim
13\,M_{\odot}$\,yr$^{-1}$), and likely the outflow as well. We
thus consider the redshifted BALs in J1035+1422, and in other
quasars of our sample in general, to be compelling evidence for the
long-sought inflows that directly feed the accretion disks and
consequently the SMBHs. Our result demonstrates that the accretion
disks in quasars are supplied with external gas transported inward
from a distant inventory, at least in some objects.

Interestingly enough, the outer distance range of the inflow $3.2~{\rm pc} 
\simeq 3.3 \times 10^{3}\,R_{\rm g}$ is comparable to the radius of
the postulated dusty torus, which was evaluated by the observed
luminosity as the dust sublimation distance $R_{\rm sub} \approx 3.0
\simeq 3.1 \times 10^{3}\,R_{\rm g}$\cite{Netzer2015}. It is likely
that the inflow headstreams set off from the dusty torus, a natural
reservoir of gas supply within the sphere of influence of the SMBH.
Part of the inflow gases may also originate from a highly extended
disk suggested in some models\cite{CollinZahn1999} to connect the
accretion disk and the torus, which is gravitationally unstable and
clumpy. In both cases the inflow may be formed by gases drifting
inward from all the azimuthal directions, resulting in an axially
symmetric geometry for the bulk of the inflow. Several mechanisms
may be responsible for removing the angular momentum of (a tiny)
part of the gases therein, such as star-formation and stellar winds,
collisions between clumps, tidal disruption of clumps by the
SMBH\cite{Terlevich1996,Wada2003,JM2017}. The resulting gases with
reduced angular momentum will spiral in via sub-Keplerian motion,
among which those having the least angular momentum are accelerated
to the fastest speeds as observed. The consideration of the residual
angular momentum also favors an axial symmetric geometry for the
bulk of the inflows. This raises an interesting postulate that the
existence of a dusty torus in galactic nuclei, and the processes and
efficiency to redistribute the specific angular momentum for gases
therein, are essential for triggering, and determining the level of,
quasar activity.

One important question to address is the ubiquity of the
disk-feeding inflow in quasars. If the inflows are aligned at large
inclination angles with respect to the axis of the general plane
(e.g. in the case of originating from the torus; as shown in
Figure\,\ref{Figure4_cartoon}), the vast majority of the LOS to the
central radiation source that intersect the inflows are expected to
be heavily obscured by the dusty torus. This makes the inflow
essentially undetectable, except for those LOS passing at a grazing
incidence angle to the upper surface of the torus, or passing
through in-between clumpy clouds. The eight objects found with
redshifted H$_{n+2}$ and He\,I$^{*}_{n}$ BALs out of $10^{5}$ quasars
may just be such extremely rare cases. This is supported by the fact
that the optical-ultraviolet spectra of the eight objects are
heavily reddened. Considering the difficulty in observation, the
disk-feeding inflows as discovered here may actually be much more
common in quasars than they appear to be, and should be considered
as an indispensable component---the last piece of the puzzle that
falls into place---for the paradigm of quasar black hole accretion.

Near the boundary of the inflow, gases with low (column) densities
may disperse gradually, forming comet-like tails as spiraling in. If
the tails get too dispersed to be shielded, they may be stopped and
even reversed by the strong quasar radiation field, forming new (or
becoming part of existing) shielding outflows. They may reveal
themselves in blueshifted BALs of resonant metal ions, which can be
more easily and commonly observed than redshifted H$_{n+2}$ and
He\,I$^*_{n}$ BALs\cite{Hall2013}. It is worth noting that both the
location and physical conditions of the redshifted and blueshifted
H$_{n+2}$ and He\,I$^*_{n}$ BALs are similar to those of the BELR in
quasars. We found in our preliminary photo-ionization model
calculations that the overall H$\alpha$ emission lines produced by
the disk feeding inflow and the BAL outflow are comparable in
strength to the BELs observed in quasars (Table\,1 and Extended Data
Fig.\,\ref{Fig11_PreHalpha} in \textbf{Method}). It is tempting to
hypothesize that the BELR is nothing special but merely a
combination of the inflow, outflow, and accretion disk in quasars.

\begin{addendum}
 \item 
This is a pre-print of an article published in Nature. The final authenticated version is available online at: https://doi.org/10.1038/s41586-019-1510-y.

The authors acknowledge the use of the Hale 200-inch Telescope at
Palomar Observatory through the Telescope Access Program (TAP) that
made these observations possible. Funding for SDSS-III has been
provided by the Alfred P. Sloan Foundation, the Participating
Institutions, the National Science Foundation, and the U.S.
Department of Energy Office of Science. The SDSS-III Web site is
http:// www.sdss3.org/. H.Z., L.S. and X.P. acknowledge support by
the National Natural Science Foundation of China (NSFC grant No.
11473025) and by the SOC program (CHINARE2017-02-03).

 \item[Author contributions]
H.Z. conceived the project. X.P. led the data acquisition and data
reduction. X.S. analyzed the data and performed the photo-ionization
simulations. W.Y., X.S., L.H., and G.L. contributed to writing the
manuscript. All coauthors provided critical feedback to the text and
helped shape the manuscript.

\item[Author information]
The authors declare no competing financial interests. Correspondence
and requests for materials should be addressed to H.Z.
(zhouhongyan@pric.org.cn).

% \item[Reviewer Information] .

\end{addendum}

\clearpage

\section*{References}
%\bibliography{lyacovfrac,lyacovfrac_add1}

\clearpage
%%%%%%%%%%%%%%%%%%

\renewcommand{\figurename}{\textbf{Fig.}}

\begin{figure}
\begin{center}
\includegraphics[width=0.8\textwidth]{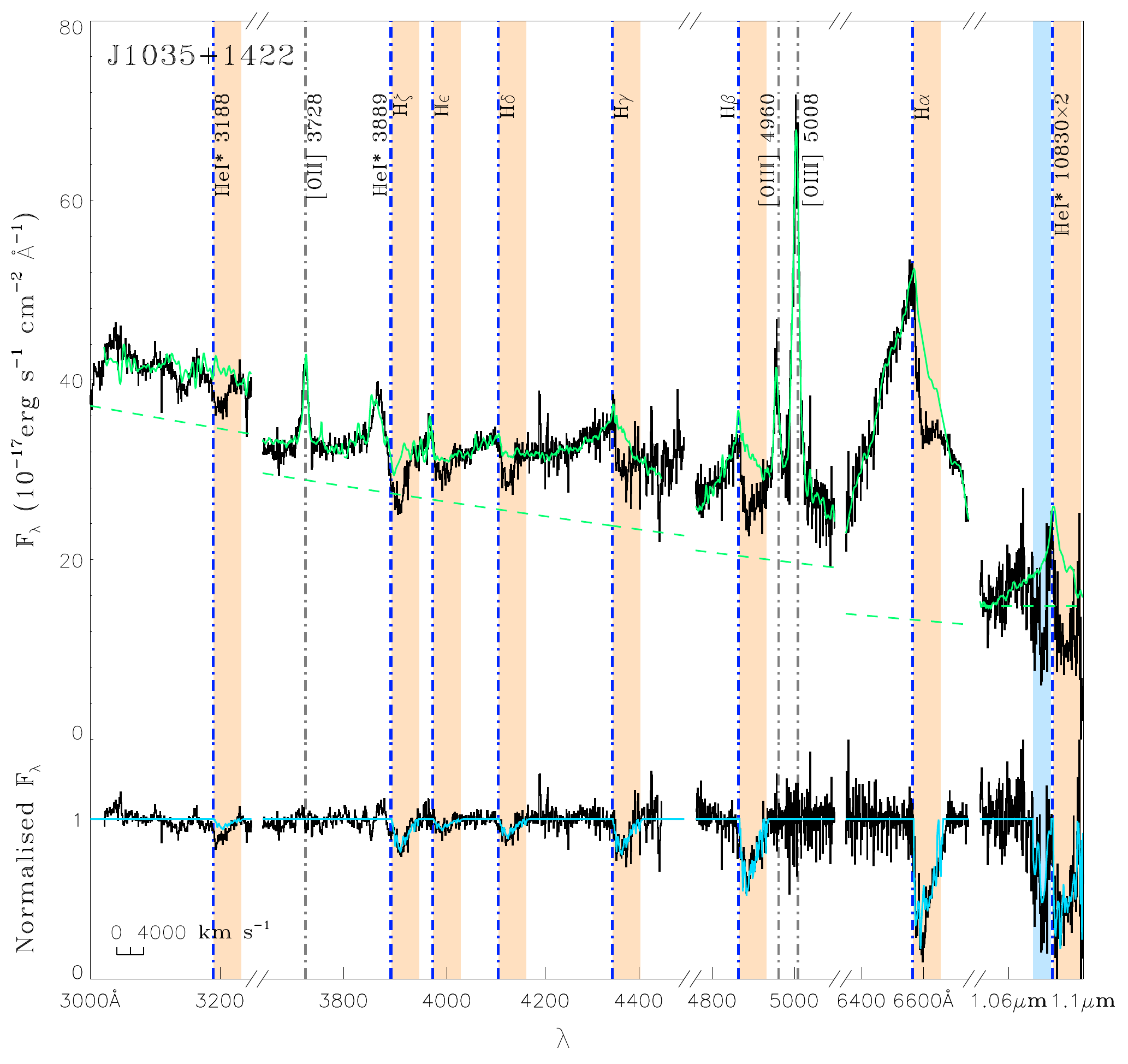}
\end{center}
\caption[]{\textbf{Broad absorption lines (BALs) of the hydrogen
Balmer series H$_{n+2}$ and the meta-stable neutral helium
multiplets He\,I$^{*}_{n}$ revealed in the rest-frame spectrum, from
near-ultraviolet to near-infrared, of the quasar J1035+1422.} In the
upper part, the black line shows the observed flux, the green solid
line represents the unabsorbed quasar template spectrum obtained by
the `pair-matching' method (see \textbf{Method}), and the green
dashed lines show the underlying power-law continuum. In the lower
part, the black line shows the normalized BAL spectrum. The
redshifted BALs are normalized to the continuum only (with the
modeled emission line flux subtracted), since they are found to be
not obscuring the broad emission line region. The blueshifted He\,I*
$\lambda 10830$ BAL is, in contrast, normalized using the total flux
of the modeled template, as it is assumed to be fully obscuring both
the continuum source and the broad emission line region. The cyan
line represents the best-fit model BALs, which are employed to
derive ion column densities from the H$_{n+2}$ and HeI$^{*}_{n}$
lines (The velocity profiles of the optical depth and covering
factor for the H$\alpha$ BAL are shown in Extended Data Fig.
\ref{Fig5_Halpha} of \textbf{Method}). The blue dotted-dashed lines
mark the rest wavelengths for the H$_{n+2}$ and HeI$^{*}_{n}$ transitions.
The quasar systemic redshift is determined from narrow emission
lines, including [OII] (marked as gray dotted-dashed vertical
lines).} \label{Figure1_spectrum}
\end{figure}

\clearpage

\begin{figure}
\begin{center}
\includegraphics[width=0.55\textwidth]{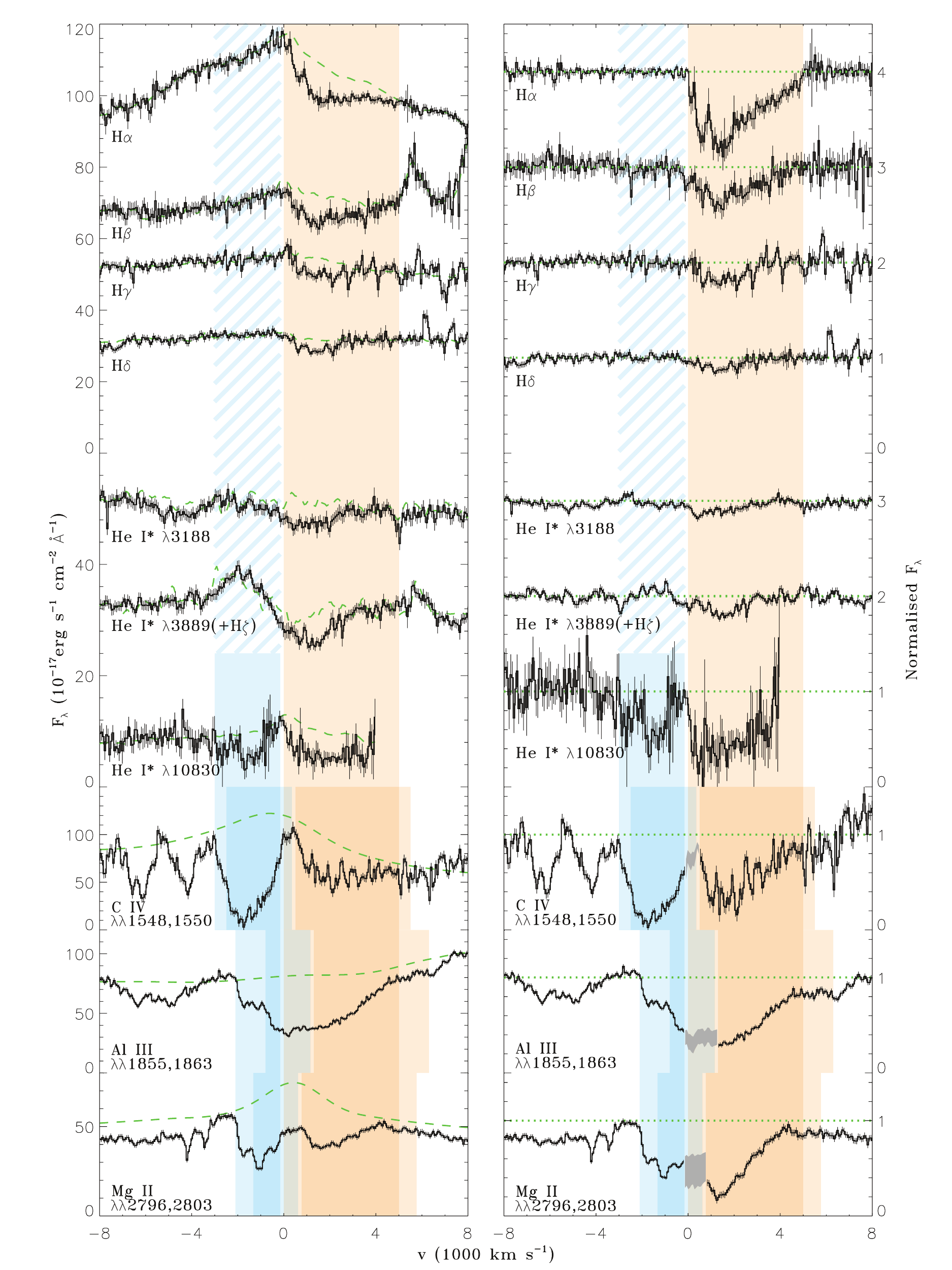}
\end{center}
\caption[]{\textbf{Close-up of the observed (left) and normalized
(right) BAL spectra of J1035+1422 in selected H Balmer (H$\alpha$ --
H$\delta$) and meta-stable He\,I* (He\,I*$\lambda 10830$, He\,I*$\lambda 3889$(+H$\zeta\lambda$3890), and He\,I*$\lambda 3189$) in
comparison with those of the metal lines, C\,IV, Al\,III, and
Mg\,II, plotted in their common velocity space.} The velocity ranges
for the redshifted and blueshifted BAL systems (colored regions in
red and blue respectively) are determined according to the
normalized H$\alpha$ and He\,I*$\lambda 10830$ BAL spectra. The
seeming velocity difference between H, He, and metal BALs are mostly
due to the fact that all of the metal lines are actually doublets
with a velocity offset of 500, 1,300, and 770 km~s$^{-1}$ between
the member lines of C~IV, Al~III, and Mg~II, respectively. The
best-fit quasar composite spectrum is shown by the green dashed
lines. It is obvious that the redshifted C\,IV trough is shallower
than the blueshifted C\,IV trough and then the redshifted Mg\,II and
H$\alpha$ troughs. The He\,I*$\lambda 3889$ line is severely blended
with H$\zeta\lambda$3890, with a measured optical depth ratio of
$0.7/0.3$. Note that all of the H and He I* BALs show solely
redshifted absorption troughs except for He\,I*10830, the
overwhelmingly strongest in strength of the multiplets. This implies
drastically distinctive physical condition of the redshifted BAL gas
from that of the blueshifted one. Unlike emission lines, which may
be complicated by various effects, such as obscuration and
projection, redshifted absorption lines are clean in kinematics and
are of a robust indication for inward motion of the absorbing gas
along the LOS\cite{Drew1995}. } \label{Figure2_BALs}
\end{figure}

\clearpage

\begin{figure}
\begin{center}
\includegraphics[width=\textwidth]{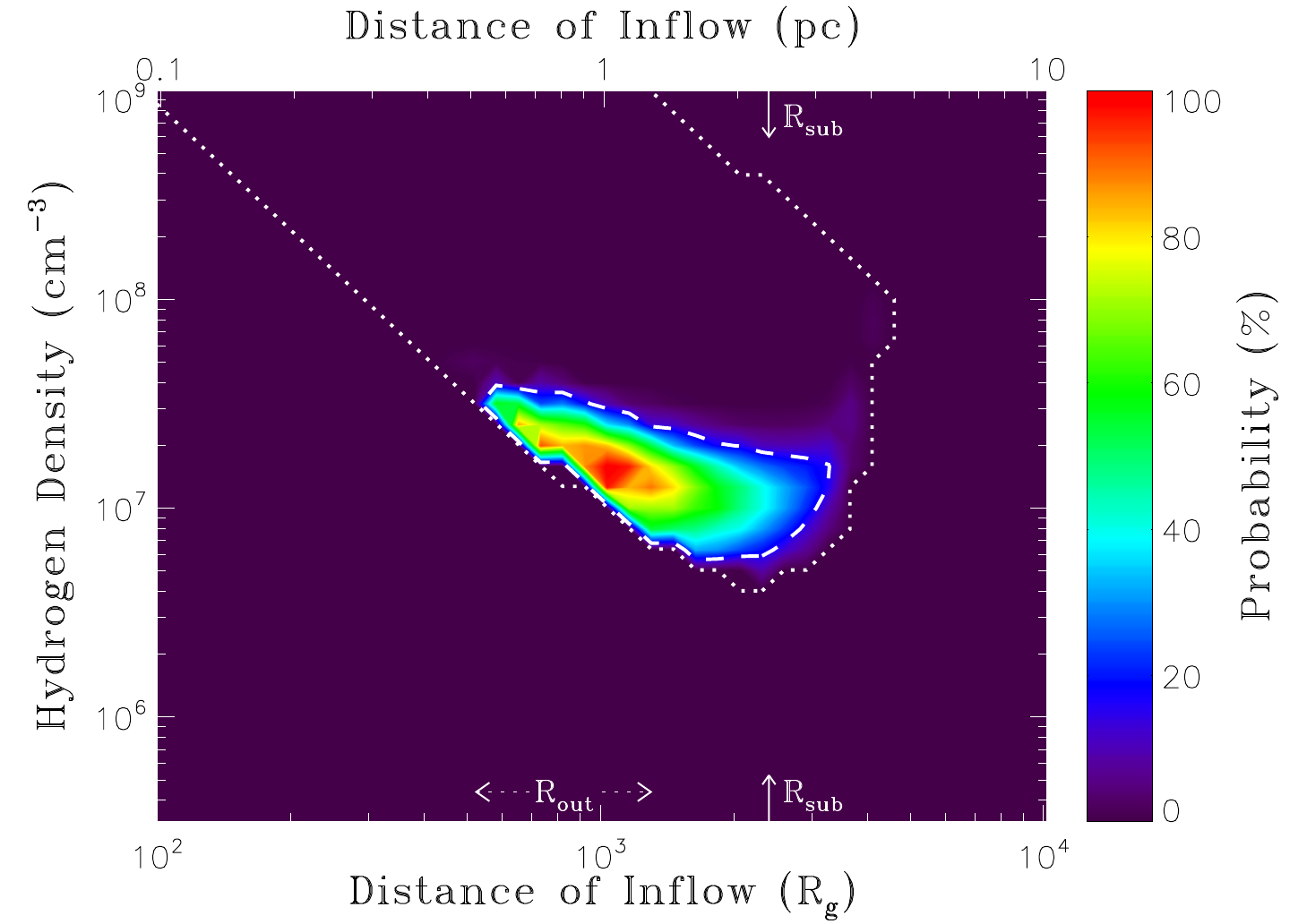}
\end{center}
\caption[]{\textbf{Normalized probability density distribution in
the parameter space of the total hydrogen density $n_{\rm H}$ and the
characteristic distance $R_{\rm IF}$ of the inflow from the central
black hole in J1035+1422.} This is obtained by matching the
photo-ionization models with the observed redshifted BALs of H\,I
Balmer, He\,I*, C\,IV, and UV Fe\,II. The best-fit values are
$n_{\rm H}\approx 10^{6.75}-10^{7.60}~{\rm cm}^{-3}$ and $R_{\rm IF}\approx 0.514
- 3.171~{\rm pc}$ (or $525-3,236~R_{\rm g}$) at the 90\%~confidence level.
Over-plotted are contours of the total hydrogen column density of
the inflow models, indicating a well constrained range of
$N_{\rm H}\approx 10^{23.22}-10^{23.79}~{\rm cm}^{-2}$ (peaked at
$N_{\rm H} \approx 10^{23.46}~{\rm cm}^{-2}$). The dotted
arrows mark a postulated range of the outer accretion disk radius,
which is taken as the previously reported disk radii derived from
observations for a number of quasars with similar estimated
self-gravity radii to that of J1035+1422, from 525 $R_{\rm g}$ (3C332) to
1,300 $R_{\rm g}$ (3C390.3). The dotted line shows the parameter envelope
(90\%) inferred by models without considering UV Fe\,II BALs. The
inner surface of the dusty torus derived as the sublimation radius
$R_{\rm sub}$ is indicated by the solid arrows for comparison. These
results suggest that the inflow gas is dense and thick, and is
located largely in between the accretion disk and the presumed dusty
torus as sketched in Fig. \ref{Figure4_cartoon}.}
\label{Figure3_probability}
\end{figure}

\clearpage

\begin{figure}
\begin{center}
\includegraphics[width=\textwidth]{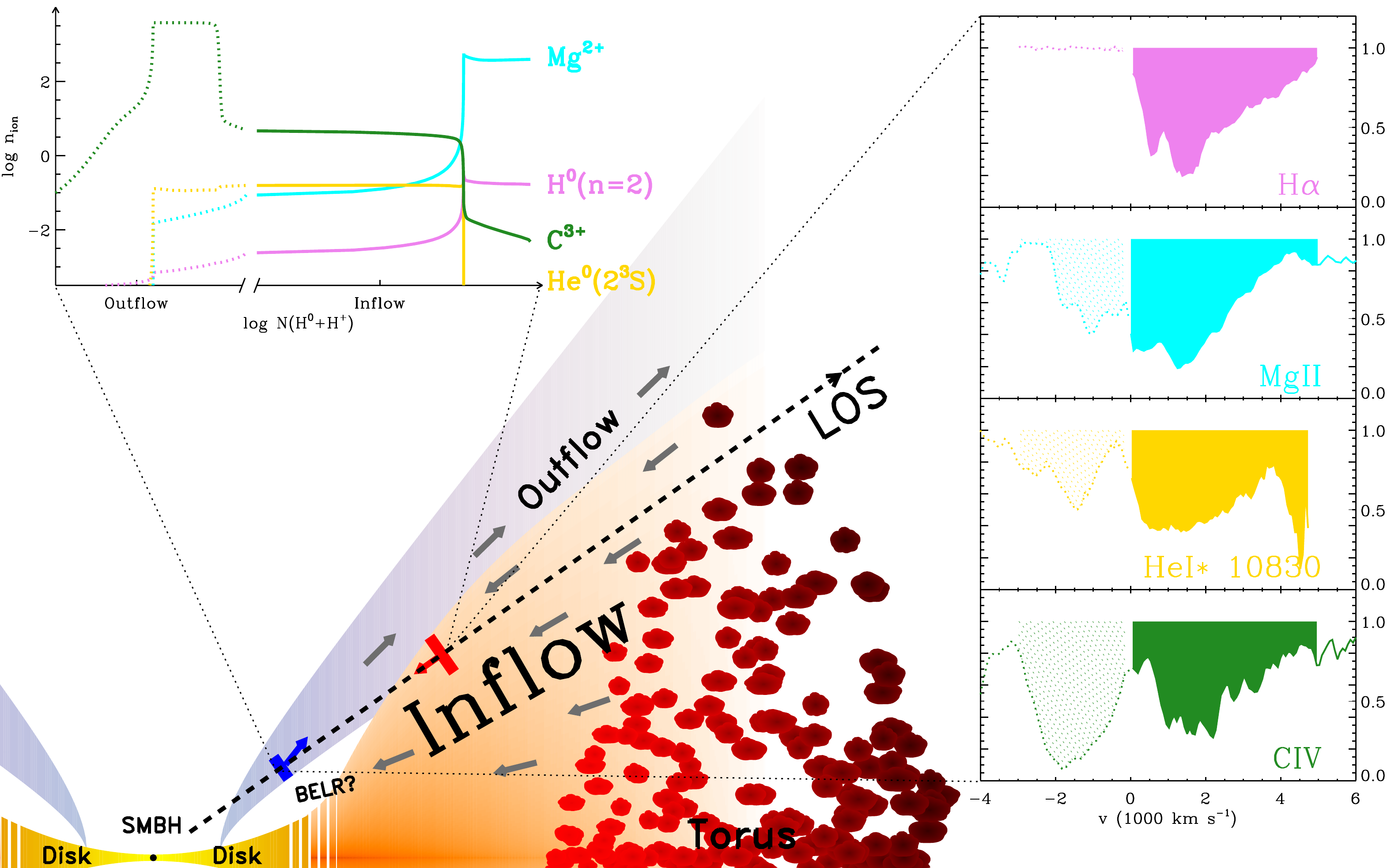}
\end{center}
\caption[]{\textbf{Schematic view of the central engine of quasars
with an inflow revealed by the hydrogen H$_{n+2}$ and helium
He\,I$^{*}_{n}$ BALs observed in J1035+1422 and its analogues.} The
inflow is positioned to be at parsec scale (about a thousand
gravitational radii $R_{\mathrm{g}}$ of the black hole), located
in-between an accretion disk and a dusty torus. The in-falling
velocities of the gases along the line-of-sight (LOS) span from zero
to as high as $5,000~{\rm km~s}^{-1}$ (about the free-fall speed just
beyond the outer radius of the accretion disk), indicating that the
flow is being accelerated under the pull of the central SMBH. A
large amount of cold, dense and high-column density gas as inferred
from the BAL modeling provides direct and sufficient mass supply to
feed the accretion disk. The inflow originates most likely from the
torus, formed by materials therein that lose a substantial fraction
of their angular momentum via various processes and hence fall
inward. Viewed at a modest inclination angle, the LOS may intercept
both the inflow and outflow, leaving imprints as both redshifted and
blueshifted BALs that are observed in J1035+1422 (shown in the right
inset are four representative BALs in their common velocity space).
Color codes for the BALs: H$\alpha$ in magenta, Mg~II in cyan,
He~I*$\lambda10830$ in yellow, and C~IV in olive. The densities of
the corresponding ions as a function of the depth along the LOS
(represented by the total hydrogen column density $N_{\rm H^{0}+H^{+}}$)
are also shown in the upper-left inset with the same color code.}
\label{Figure4_cartoon}
\end{figure}

%%%%%%%%%%%%%%%%%%%%%%%%%%%%%%%%%%%%%%%%%%%%%%%%%%%%%%%%%%%%%%%%%%%%%%

\clearpage

%%%%%%%%%%%%%%%%%%%%%%%%%%%%%%%%%%%%%%%%%%%%%%%%%%%%%%%%%%%%%%%%%%%%%%

\begin{methods}

\section{Diagnostic Sensitivity of the H~I Balmer and He~I * BALs within the Torus Scales.}

The absorption lines in H~I Balmer series and the meta-stable He~I
multiplets cover significantly wide ranges in both wavelengths and
oscillator strengths. These advantages make them good probes to the
physical properties of gaseous medium, much more powerful than the
conventional resonant metal absorptions, such as C~IV, Si~IV or
Mg~II. We can measure the residual flux $f_{\rm r}(\lambda)$ on an
observed absorption-line spectrum. Assuming that the absorption-free
flux is $f_{0}(\lambda)$, we have
$f_{\rm r}(\lambda)=f_{0}(\lambda)\cdot e^{-\tau(\lambda)}$, where
$\tau(\lambda)$ is the optical depth determined by how much
absorbing material there is along the LOS. The latter is often
quantitatively described as the column density of the absorbing
material. The optical depth at a given point in an absorption
profile $\tau(\lambda)$ can often be well determined in a spectrum
with good data quality when it is within the range of 0.05--3 (the
corresponding normalized residual flux $e^{-\tau(\lambda)}$ is
0.95--0.05 in the absorption trough, meaning that the line is
neither too weak nor severely saturated). Thus for H$\alpha$ at
$6564~\mathrm{\AA}$ the measurable range of column density is
$2.0\times
10^{11}<\frac{dN_{\mathrm{col}}(\mathrm{H}^0_{n=2})}{d\lambda}(\mathrm{cm}^{-2}\mathrm{\AA}^{-1})<1.2\times
10^{13}$, while for H$\kappa$ at $3750~\mathrm{\AA}$ the measurable
range is $1.9\times
10^{14}<\frac{dN_{\mathrm{col}}(\mathrm{H}^0_{n=2})}{d\lambda}(\mathrm{cm}^{-2}\mathrm{\AA}^{-1})<1.1\times
10^{16}$. Therefore, the Balmer series enables measurement of the
column density over $\sim 5$ orders of magnitude. The He I*
mutiplets (e.g. He I* $\lambda 2829$ to He I* $\lambda 10830$) also
covers a range of $8.9\times
10^{10}<\frac{dN_{\mathrm{col}}(\mathrm{He}^0
2^3\mathrm{S})}{d\lambda}(\mathrm{cm}^{-2}\mathrm{\AA}^{-1})<6.8\times
10^{15}$ or even wider. In contrast, the C~IV $\lambda\lambda
1548,1550$ doublet only covers a range of $1.2\times
10^{13}<\frac{dN_{\mathrm{col}}(\mathrm{C}^{3+}_{\mathrm{ground}})}{d\lambda}(\mathrm{cm}^{-2}\mathrm{\AA}^{-1})<1.5\times
10^{15}$.

In Extended Data Fig. \ref{Fig1_Hydrogen} and Fig.
\ref{Fig2_Helium}, we show a simple demonstration of the diagnostic
sensitivity of H~I Balmer and He~I* absorption lines. We assume that
the central SMBH has a mass of $M_{\bullet}=10^9~M_{\mathrm{\odot}}$
and an Eddington ratio of 0.1. We place a slab of homogeneous
absorbing gas with solar abundances at a distance $d_{\mathrm{abs}}$
to the central SMBH that varies from $10^2$ to $10^4$ gravitational
radius $R_{\mathrm{g}}$ ($\equiv\frac{GM_{\bullet}}{c^2}$, i.e.
roughly where the outflow is supposed to originate before reaching
the dusty torus). The ionization (and the ionic column densities at
the levels of interest) of the modeled gas is evaluated with the
photo-ionization code CLOUDY (the latest version)\cite{Ferland1998}.
Given $N_{\mathrm{col}}(\mathrm{H}^0_{n=2})$ and
$N_{\mathrm{col}}(\mathrm{He}^0 2^3\mathrm{S})$ as functions of
$d_{\mathrm{abs}}$, the density $n_{\mathrm{H}}$ and total column
density $N_{\mathrm{H}}$ of the gas, and assuming a Gaussian
velocity dispersion corresponding to a full width at half maximum
(FWHM) of $3,000~\mathrm{km~s}^{-1}$, we can estimate the optical
depths at line centers $\tau_{\mathrm{center}}$ for various
transitions. As long as $\tau_{\mathrm{center}}$ is between 0.05 and
3, we consider the method sensitive for measuring the corresponding
ionic column densities.

The colored area in the plots shows the sensitive range for each
individual line. For $N_{\mathrm{H}}=10^{20}~\mathrm{cm}^{-2}$, only
denser and farther-away medium has enough neutral gas to render
considerable optical depths in H$\alpha$ and He I* $\lambda 10830$
lines. When $N_{\mathrm{H}}$ increases,
$N_{\mathrm{col}}(\mathrm{H}^0_{n=2})$ and
$N_{\mathrm{col}}(\mathrm{He}^0 2^3\mathrm{S})$ also increase. At a
certain point, these ionic column densities are so large in gas with
intermediate density that H$\alpha$ and He I* $\lambda 10830$ become
saturated. While higher-order lines with smaller oscillator
strengths (e.g. H$\delta$ and He I* $\lambda 3889$) become sensitive
probes, instead. Therefore, by including the whole series of H~I
Balmer and HeI$^{*}$ lines, we can reliably measure
$N_{\mathrm{col}}(\mathrm{H}^0_{n=2})$ and
$N_{\mathrm{col}}(\mathrm{He}^0 2^3\mathrm{S})$ in the vast majority
of the parameter space of our interest, especially for thick medium
($N_{\mathrm{H}}>10^{22}~\mathrm{cm}^{-2}$).

\section{Redshifted H I Balmer/He I* BAL and Mini-BAL Systems in Quasars.}

Through a systematic search for H~I Balmer and He I* absorption
systems in the SDSS quasar catalog ($\sim 10^5$ quasars with $z <
1.3$), we find $\sim 50$ such systems with He I* $\lambda 3889$
absorption and at least three Balmer absorption lines detected at $>
3\sigma$ significance. NAL, mini-BAL, and BAL systems account for
$\sim 1/3$ of this sample, respectively. $\sim 1/2$ show net
blueshited troughs, $\sim 1/4$ show net redshifted troughs, while in
the rest $\sim 1/4$ the troughs are either unshifted or extend both
bluewards and redwards. Of great interest are those systems with
redshifted BAL or mini-BAL troughs, which are candidates of inflows
proximate to their central engines. In Extended Data Fig.
\ref{Fig3_sample}, we present seven additional objects manifesting
redshifted BALs or mini-BALs for H~I Balmer and He I* absorptions,
besides J1035+1422. One of the mini-BAL systems, the H~I Balmer and
He I* absorption system in J1125+0029, has been scrutinized and
confirmed to be a parsec-scale inflow\cite{Shi2016}, even though we
could not confirm if the inflow can go further into the center to
directly feed the accretion disk around the the central SMBH. This
is because that the H~I Balmer and He~I* absorption only trace the
intermediate- to low-ionization and neutral regions in the absorbing
medium. Since the SDSS spectra of these objects cover wavelengths no
shorter than $1,700~\mathrm{\AA}$ in their rest frame, the
high-ionization lines (C~IV, Si~IV, and N~V doublets, etc.) are all
missed out. Hence, we know little about the highly ionized region in
the absorbing medium, and the potential relation between the
redshifted Balmer/He I* BAL or mini-BAL systems and the most common
HiBAL (high-ionization BAL) systems (whether redshifted or
blueshifted).

J1035+1422 is also listed in our catalog of redshifted H~I Balmer/He
I* BAL quasars, showing the largest width of absorption troughs
therein. The multiple archival and our follow-up spectroscopic
observations have ensured the wavelength coverage from C~IV
$\lambda\lambda 1548,1550$ to He~I*$\lambda 10830$. Therefore, both
high- and low-ionization lines can be measured, and the absorbing
gas along the line of sight could thus be explored further. In the
following sections, we present a detailed analysis on the BAL
systems in J1035+1422 and characterize their roles in the
theoretical scheme of the quasar structure.

\section{Optical and NIR Spectroscopic Observations, and Data Reduction.}

The BOSS spectrum for J1035+1422 was observed on March 24, 2012 with
an exposure time of $4,500~\mathrm{s}$. We conducted a follow-up
campaign aiming at this object with the intermediate-resolution
DoubleSpec and TripleSpec spectrograph equipped on the Palomar
200-inch Hale telescope in the optical and NIR. The DoubleSpec data
were obtained on Apr 23, 2014 with a total exposure of $4\times
500~\mathrm{s}$, and the TripleSpec data were acquired on Mar 13,
2017 with a total exposure of $12\times 120~\mathrm{s}$. An archival
broad-band optical-to-NIR spectrum is also available, which was
obtained by VLT/X-shooter on Dec 6, 2014 through the ESO program
094.A-0087(A), with a spectral resolution of $R\equiv \lambda/\Delta
\lambda \sim 10,000$. The total exposure time is $\sim$ 40 min. We
follow the standard procedure in reducing the raw data, using
IRAF\footnote{http://iraf.noao.edu/} and the IDL SpexTool
package\cite{Cushing2004}. The
Catalina\footnote{http://nesssi.cacr.caltech.edu/DataRelease/}
monitoring of J1035+1422 shows negligible variability in V-band
during a period of 8 years (2005 Apr 9 to 2013 Oct 26). Thus, We
combine the relatively low resolution BOSS and DoubleSpec/TripleSpec
data and use the combined spectrum to calibrate the X-shooter
{\'e}chelle spectrum. The BAL troughs are clearly resolved in the
X-shooter spectrum.

\section{The Central Engine.}

Though the rest-frame UV through optical to NIR color of J1035+1422
is the bluest in the sample of quasars with redshifted H~I Balmer
and He~I* absorption lines, it is obviously redder than normal
optically-selected quasars. Using the SDSS quasar composite
spectrum\cite{VandenBerk2001} reddened with the SMC-type extinction
curve\cite{Gordon2003} to match the flux in the absorption-free
windows in the rest-frame UV band, we find $E(B-V)\approx 0.12$,
implying a significant amount of dust reddening.

We estimate the mass of the central SMBH using the H$\alpha$ broad
emission line width and the rest-frame optical luminosity, and
adopting the radius-luminosity scaling relation\cite{GreeneHo2005}.
The monochromatic luminosity at 5100~\AA~in the quasar's rest-frame,
$L_{5100}$, is derived using the measured
$f_{\lambda}(5100~\mathrm{\AA})$ and assuming a cosmology with
$H_0=70~\mathrm{km~s}^{-1}~\mathrm{Mpc}^{-1}$,
$\Omega_{\mathrm{M}}=0.3$ and $\Omega_{\mathrm{\Lambda}}=0.7$.
Compared to Mg II and H$\beta$ emission lines, the H$\alpha$ peak is
less affected by the absorption trough. Therefore, we use the width
of the broad H$\alpha$ emission line ($\mathrm{FWHM} \approx 1.4\times
10^4~\mathrm{km~s}^{-1}$) to estimate the black hole mass, and yield
$M_{\mathrm{BH}}\approx 2.0\times 10^{10}~M_{\mathrm{\odot}}$.
Adopting a bolometric luminosity of $L_{\mathrm{bol}}=(8.1\pm
0.4)~L_{5100}$\cite{Runnoe2012} and assuming an accretion efficiency
of 0.1, we estimate the mass accretion rate to be
$\dot{M}_{\bullet}\approx
13~M_{\mathrm{\odot}}~\mathrm{yr}^{-1}$.

The sizes of quasar accretion disks are difficult to determine, and
they are generally thought to be of the order from several hundred
to a few thousand gravitational radii. Theoretical models assume the
self-gravity radius $R_{\rm sg}\simeq 6.04 \times 10^{4}
\alpha^{2/9} \eta^{-4/9} M_{\bullet}^{-2/9} (L_{\rm bol}/L_{\rm Edd})^{4/9}$
to be a natural outer disk boundary, where the vertical component of
the central gravity is balanced by the self-gravity of the disk, and
$\alpha \sim 0.01-1$ is the viscosity parameter and $\eta \equiv
L_{\rm bol}/\dot{M}_{\bullet}c^{2} \sim 5.7-42\%$ the accretion
efficiency\cite{Netzer2015}. Beyond this boundary the disk becomes
gravitationally unstable and starts to fragment into clumps, forming
gaps that prevent efficient mass accretion. We estimated a
self-gravity radius $R_{\rm sg}\approx 140\,R_{\rm g}$ in J1035+1422
for $\alpha = 0.3$ and $\eta = 10\%$. Observationally, $R_{\rm out}$
can be well modeled in a few percent of quasars with double-peaked
emission line profiles. For some of those with the estimated
self-gravity radii similar to that of J1035+1422 within a factor of
two (namely 3C 332, Pictor A, 3C 17, Arp 102B, and 3C 390.3), their
modeled outer disk radii as available from the literature
\cite{EracleousHalpern1994,EracleousHalpern2003,WuLiu2004,LewisEracleous2006}
fall into $R_{\rm out}\sim 540-1,300\,R_{\rm g}$. We thus consider
the outer disk radius of J1035+1422 to lie somewhere from a few
hundred to about one thousand gravitational radii.

\section{The BAL Systems.}

The redshifted BAL troughs can be easily identified for various
transitions, including the H~I Balmer series, the meta-stable
He~I$^*$$\lambda\lambda 3,188,3,889,10,830$, C IV$\lambda\lambda
1,548,1,550$, Al III$\lambda\lambda 1,854,1,862$, and Mg
II$\lambda\lambda 2,796,2,803$. For He I*$\lambda 10830$, C IV, and
Mg II, separate, blueshifted troughs are also clearly observed,
while for Al III doublets with a larger wavelength interval between
its two transitions, the blueshifted BAL trough only extends
bluewards smoothly. For Balmer lines and other weaker He I* lines,
no evident blueshifted BAL absorption has been detected.
Furthermore, comparing the spectra with the SDSS quasar composite,
we find that considerable rest-frame UV flux is absorbed by Fe II
between rest-frame 2,000 and $2,800~\mathrm{\AA}$, and wavelengths
straddling rest-frame $1,600~\mathrm{\AA}$. Due to the numerous line
transitions at these wavelengths, the absorption troughs overlap
seriously with each other, rendering it impossible to identify the
profile of any individual BALs.

\section{Measurement of the Redshifted BALs.}

The redshifted H~I Balmer BALs, from H$\alpha$ to H$\epsilon$ can be
readily identified on the observed spectrum of J1035+1422 with
absorbing troughs within the velocity range from $\sim 0$ to $\sim
5,000~\mathrm{km~s}^{-1}$ with respect to the quasar's rest-frame
accurately determined by narrow emission-lines, including
[O~II]$\lambda$3727 and [O~III]$\lambda\lambda$4959,5007. Due to its
relative weakness ($\lambda f_{ij}$ ratio of 2.42 between
H$\epsilon$$\equiv$H$_{7}$ and H$\eta$$\equiv$H$_{9}$), redshifted
H$\eta$ BAL is only marginally detected with a similar velocity
structure. H$\zeta$$\equiv$H$_{8}\lambda3890$ should also be
present, but it is heavily blended with He~I$^*\lambda3889$. We use
the pair-matching method to recover the absorption-free
spectrum\cite{Zhang2014,Liu2015} of J1035+1422. The underlying
assumption is that if the spectrum of a non-BAL quasar resembles the
spectrum of a given BAL quasar in the absorption-free portions, they
are intrinsically similar, and the non-BAL quasar then provides a
good approximation to the unabsorbed flux striding over the BAL
troughs. We choose non-BAL quasars from the BOSS quasar catalog, and
fit the spectra with that of J1035+1422. The wavelength ranges
affected by BAL troughs ($v\sim 0-5,000~{\rm km~s}^{-1}$ for the above
mentioned Balmer lines) are masked out during the fit. If the
reduced $\chi^2 < 1.5$, we consider it an acceptable match. The mean
spectrum of all accepted non-BAL quasar spectra is used as the
unabsorbed template, and the variance is used to estimate the
uncertainty of the template.

From the spectra normalized using the template, we can derive the
covering factor and the optical depth for the redshifted Balmer BALs
as a function of velocity shift $v$ following

\parbox{10cm}{\begin{eqnarray*} I_{\mathrm{H}\alpha}(v) &=& [1-C_{\mathrm{f}}(v)]+C_{\mathrm{f}}(v)e^{-\tau_{\mathrm{H}\alpha}(v)} \\ I_{\mathrm{H}_i}(v) &=& [1-C_{\mathrm{f}}(v)]+C_{\mathrm{f}}(v)e^{-\tau_{\mathrm{H}\alpha}(v)\lambda_{\mathrm{H}_i}f_{\mathrm{H}_i}/\lambda_{\mathrm{H}\alpha}f_{\mathrm{H}\alpha}} \end{eqnarray*}} \hfill
\parbox{1cm}{\begin{eqnarray}\label{eq1}\end{eqnarray}}

where $I(v)$ is the normalized flux, $C_{\mathrm{f}}(v)$ is the
covering factor, and $\tau(v)$ is the optical depth.

In practice, because of the relatively broad widths of the BAL
troughs and the diversity of the profiles that the emission lines
may have, the pair-matching and the profile extraction are performed
iteratively: an initial guess for the red wings of the emission
peaks is used as a rough estimate for the BAL profiles. Based on the
absorption corrected spectra derived by employing these profiles, a
more realistic template can be further achieved. This procedure
continues until a set of self-consistent emission template and
absorption profiles are finally obtained.

The resultant template for H~I Balmer emissions is plotted in Figure
\ref{Figure1_spectrum} in the main text. Even if based on a
premature template, we can see that the depth of the apparently
unsaturated H$\alpha$ trough is not significantly larger than
H$\beta$. This is hard to explain considering the large $\lambda
f_{ij}$ ratio of 7.26 between H$\alpha$ and H$\beta$, unless we
assume the absorbing medium only obscures part of the background
continuum source, leaving the emission line region unobscured. In
this analysis, we adopt this assumption so that the normalized flux
used in Eq.\ref{eq1} is evaluated by removing the emission lines of
the template and dividing the residual spectrum by the continuum.
The resultant covering factor and true optical depth as a function
of velocity shift, are plotted in Extended Data Fig.
\ref{Fig5_Halpha}.

The integral $\int \tau_{\mathrm{H}\alpha}(v)\,dv$ through the BAL
trough yields a column density of
$N_{\mathrm{col}}(\mathrm{H}^0_{n=2})=1.98\pm 0.14\times
10^{15}~\mathrm{cm}^{-2}$ for the redshifted system. The column
density at the meta-stable He$^0$ level $2^3\mathrm{S}$ of the same
absorbing gas is measured using He I* $\lambda 3889$, given that the
He I* multiplets and the Balmer series BAL are found to share the
same absorption profile. (In fact, we find the normalized spectrum
of redshifted He I* $\lambda 10830$ BAL, which is deemed fully
saturated, can be well described by the profile of the covering
factor $C_{\rm f}(v)$ derived from Balmer BALs, see Fig.
\ref{Figure2_BALs}.) By removing the contribution of H$_8$
absorption from the trough at rest-frame $\sim 3,900\ \mathrm{\AA}$
and applying the fractional distribution $\tau (v)/\int \tau(v)\,dv$
extracted from Balmer series, we find
$N_{\mathrm{col}}(\mathrm{He}^0 2^3\mathrm{S})=1.11\pm 0.14\times
10^{15}~\mathrm{cm}^{-2}$.

The absorption troughs of metal lines in the range of $\sim 1,500 -
3,300~\mathrm{\AA}$ relative to the quasar's rest-frame, including C
IV, Al III, Fe II, Mg II, etc., are hard to characterized. Both the
UV Fe II and the redshifted BAL doublets of those alkaline-like
metal ions show seriously overlapping troughs. And with blueshifted
absorptions present for at least C IV, Al III, and Mg II, the
absorption-free windows in the rest-frame UV band is too limited to
reveal the unabsorbed flux through the pair-matching method. The
initial fitting using SDSS quasar composite suggests that the
spectra around rest-frame 1,500, 1,600, 1,960, 1,980 and
$3,100~\mathrm{\AA}$ are minimally affected by absorption,
consistent with our knowledge on the wavelength distribution of Fe
II multiplets.

\section{Photo-ionization Models of the Reshifted BAL Inflow}\label{inflow model}

\subsection{The primordial model}

Since all gaseous medium is in the vicinity of the central SMBH, the
ionization state of the inflowing gas giving rise to the redshifted
absorption system is dominated by the radiation from the central
engine. To estimate the physical conditions, we employ CLOUDY to
simulate the physical processes in the medium. The simplest model is
a slab of gas with a uniform density and chemical composition,
irradiated directly by the central continuum source. Such primordial
models can be fully described using parameters including
$n_{\mathrm{H}}$, $N_{\mathrm{H}}$, ionization parameter at the
inner (illuminated) surface $U$, abundance of elements, and the
spectral energy distribution (SED) of the incident radiation. The
CLOUDY models use these parameters and output the column densities
of various ions observed in absorption lines. Comparing the output
with our measurements, we can assess which values of the model
parameters are reasonable.

A typical AGN continuum is employed as the incident SED, which is a
combination of a UV bump described as
$\nu^{\alpha_{\mathrm{UV}}}\mathrm{exp}(-h\nu
/kT_{\mathrm{BB}})\mathrm{exp}(-kT_{\mathrm{IR}}/h\nu)$ and power
law $a\nu^{\alpha_{\mathrm{X}}}$. The UV bump is parameterized by a
UV power-law index of $\alpha_{\mathrm{UV}}=-0.5$, and an
exponentially cut-off with a temperature $T_{\mathrm{BB}}=1.5\times
10^5\ \mathrm{K}$ at high energy and
$kT_{\mathrm{IR}}=0.01\mathrm{Ryd}$ in the infrared. The power-law
component has an index $\alpha_{\mathrm{X}}=-2$ beyond
$100~\mathrm{keV}$, and $-1$ between $1.36~\mathrm{eV}$ and
$100~\mathrm{keV}$. The overall flux ratio of X-ray to optical is
$\alpha_{\mathrm{OX}}=-1.4$.

For the absorption gas with given $n_{\mathrm{H}}$ and $U$, the
model can give a specific value of $N_{\mathrm{H}}$ with which the
predicted $N_{\mathrm{col}}(\mathrm{H}^0_{n=2})$ and
$N_{\mathrm{col}}(\mathrm{He}^0 2^3\mathrm{S})$ best match the
measurements. The differences between the predicted values by this
optimal $N_{\mathrm{H}}$ and the measurements allow us to evaluate a
`probability density': $P\propto \prod_i
e^{-\frac{1}{2}(\frac{N_{\mathrm{col}}^{\mathrm{theo}}({\mathrm{ion}_i})-N_{\mathrm{col}}^{\mathrm{meas}}({\mathrm{ion}_i})}{\sigma(N_{\mathrm{col}}^{\mathrm{meas}}({\mathrm{ion}_i}))})^2}$,
where $i$ stands for different ions. If even for the optimal
$N_{\mathrm{H}}$ the differences are far beyond the measurement
uncertainties, the given $n_{\mathrm{H}}$ and $U$ appear rather
impossible. In Extended Data Fig. \ref{Fig6_prob} panel (a), we plot
the distribution of this `probability density' as a function of
density $n_{\mathrm{H}}$ and distance $d_{\mathrm{inflow}}$ (in
units of $R_{\mathrm{g}}$) from the central SMBH. The distance
$d_{\mathrm{inflow}}$ is derived as $\frac{L(\lambda <912)}{4\pi
d_{\mathrm{inflow}}^2}=Un_{\mathrm{H}}c\overline{E_{\mathrm{ph}}(\lambda
<912)}$, where $L(\lambda <912)$ is the ionizing luminosity of the
continuum source and $\overline{E_{\mathrm{ph}}(\lambda <912)}$ is
the average energy for all ionizing photons. The zone with high
probability density is a narrow, curved belt with
$d_{\mathrm{inflow}}$ between 35 and $4,000~R_{\mathrm{g}}$ and
$n_{\mathrm{H}}$ between $10^8$ and $10^{10.5}~\mathrm{cm}^{-3}$.

These highly probable models predict very large column densities for
high-ionization ions, e.g.,
$N_{\mathrm{col}}(\mathrm{C}^{3+}_{\mathrm{ground}})>4\times
10^{17}~\mathrm{cm}^{-2}$, and thus we would expect $\tau_{\mathrm{C~IV}}(v)>>1$ throughout the redshifted BAL trough. However, these
predictions do not seem to agree with the observation. Though the
metal lines cannot be measured as straightforwardly as either the
H~I Balmer or the He I* lines, an alternative strategy can be
employed to evaluate whether the predicted values are consistent
with the observation. Considering that C IV absorption has the same
$\tau(v)$ and $C_{\mathrm{f}}(v)$ profile as that of He I*, we
present in Extended Data Fig. \ref{Fig7_CIV} the model spectra,
where the flux absorbed by C IV is recovered for $\log
N_{\mathrm{col}}(\mathrm{C}^{3+}_{\mathrm{ground}})(\mathrm{cm}^{-2})=15,16$
and $17$. When the column density well exceeds
$10^{16}~\mathrm{cm}^{-2}$, in which case $\tau_{\mathrm{C~IV}}(v)>1$ throughout the redshifted BAL trough, the recovered C IV
emission line peak seems to behave abnormally, in contrast to the
smooth appearance and the consistence with the best-fit composite if
the column density is about $10^{16}~\mathrm{cm}^{-2}$. Accordingly,
we believe that the most possible value for $\log
N_{\mathrm{col}}(\mathrm{C}^{3+}_{\mathrm{ground}})(\mathrm{cm}^{-2})$
is around 15.8 with uncertainty of about 1.1, significantly smaller
than the prediction. (Even for very metal-poor gas of
$0.1~Z_{\mathrm{\odot}}$, primordial simulation models predict
$N_{\mathrm{col}}(\mathrm{C}^{3+}_{\mathrm{ground}})>10^{17}~\mathrm{cm}^{-2}$.)

\subsection{Post C$^{3+}$ region models}

Detailed investigations on the ionization structure of the
primordial models reveal that C$^{3+}$ ions, which are created by
ionizing photons with $h\nu >47.9~\mathrm{eV}$, tend to arise in the
region in front of the H I region. Beyond the C$^{3+}$ region, with
high energy photons exhausted, high-ionization ions (e.g. C$^{3+}$,
Si$^{3+}$, etc.) are negligible and low-ionization ions and neutral
atoms (e.g. Mg$^+$, H$^0$) become dominant (see Extended Data Fig.
\ref{Fig9_simu} panel (a)). it is therefore more reasonable to use
the region lying behind the C$^{3+}$ region (the `post-C$^{3+}$
region' for short) in the primordial models to estimate the physical
conditions and environment of the observed inflow, rather than using
the entire primordial models.

Still employing the primordial setup for photo-ionization
simulations, for a gaseous slab with given $n_{\mathrm{H}}$ and $U$,
now we try to find if there is a pair of $N_{\mathrm{H}}$ values for
gas between which the simulation predicts
$N_{\mathrm{col}}(\mathrm{H}^0_{n=2})$,
$N_{\mathrm{col}}(\mathrm{He}^0 2^3\mathrm{S})$, and
$N_{\mathrm{col}}(\mathrm{C}^{3+}_{\mathrm{ground}})$ within the
uncertainty of our measurements. If such $N_{\mathrm{H}}$ values are
found, they mark the inner and outer surfaces of the model suitable
for the inflow. One sample is illustrated in Extended Data Fig.
\ref{Fig9_simu} panel (a) for a primordial model with
$n_{\mathrm{H}}=10^7~\mathrm{cm}^{-3}$ and $U=10^{0.5}$
($d_{\mathrm{inflow}}\approx 1,500~R_{\mathrm{g}}$). The suggested
inner surface of the inflow model coincides with the furthest
extension of the C$^{3+}$ region. `Probability density' is estimated
in the same way with
$N_{\mathrm{col}}(\mathrm{C}^{3+}_{\mathrm{ground}})$ included. As
shown in Extended Data Fig. \ref{Fig6_prob} panel (b), the zone
corresponding to high `probability density' moves downwards, with
smaller $n_{\mathrm{H}}$ between $10^7$ and
$10^{10}~\mathrm{cm}^{-3}$. The `probability density' now presents a
bimodal distribution peaked at $n_{\mathrm{H}}\approx
10^7~\mathrm{cm}^{-3}$ and $d_{\mathrm{inflow}}\approx
2,000~R_{\mathrm{g}}$, and $n_{\mathrm{H}}\approx
10^{9.5}~\mathrm{cm}^{-3}$ and $d_{\mathrm{inflow}}\approx
100~R_{\mathrm{g}}$, respectively.

Further constraints can be introduced using another prominent
absorption feature, the UV Fe II multiplets. The Fe$^+$ ion presents
the largest number of levels in all metal ions abundant in
astrophysical gaseous medium. Following the strategy of recovering
absorbed flux according to the simulations' prediction as for C IV,
UV Fe II can reveal more information about the absorbing medium than
Mg II and Al III doublets, which originate from single levels and
the lines are saturated in our case. The result strongly favors a
`probability density' peak around $n_{\mathrm{H}}\approx
10^7~\mathrm{cm}^{-3}$. In Extended Data Fig. \ref{Fig8_UVFeII}, we
plot the recovered flux of UV Fe II bump between rest-frame 2,000
and $2,700~\mathrm{\AA}$, using the models of high probability shown
in Extended Data Fig. \ref{Fig6_prob} panel (b) ($\log
n_{\mathrm{H}}(\mathrm{cm}^{-3})=7,7.5,9$ and 9.5, respectively).
The lowest 371 levels of Fe$^+$ and the lowest 15 levels of Cr$^+$
and Ni$^+$ are considered. While the
$n_{\mathrm{H}}=10^{9.5}~\mathrm{cm}^{-3}$ model suggests no
contribution from UV Fe II absorption at all, the
$n_{\mathrm{H}}=10^7~\mathrm{cm}^{-3}$ model predicts saturated
absorptions for not only transitions of UV 1,2,3 multiplets at
around 2,300 and $2,600~\mathrm{\AA}$ from the ground term, but also
transitions from/between excited levels. The best-fit composite
(blue dashed curve in Extended Data Fig. \ref{Fig8_UVFeII}) is
consistent with the models with $\log
n_{\mathrm{H}}(\mathrm{cm}^{-3})\sim 7 - 7.5$. Therefore,
considering the Fe II absorption, we can definitely rule out the
models around the upper-left `probability density' peak at $\log
n_{\mathrm{H}}(\mathrm{cm}^{-3})=9.5$ in Extended Data Fig.
\ref{Fig6_prob} panel (b). The distribution of the updated
`probability density' $P'=P\times P_{\mathrm{UV~Fe~II}}$ is plotted in
Figure \ref{Figure3_probability} in the main text, where
$P_{\mathrm{UVFeII}}$ is estimated using the $\chi^2$ of the model
corrected for UV Fe II bump. The models of high `probability
density' is now restricted to a small area around $\log
n_{\mathrm{H}}(\mathrm{cm}^{-3})\approx 7.1$ and
$d_{\mathrm{inflow}}\approx 1,100~R_{\mathrm{g}}$.

\subsection{Radiative pressure on the inflow}

The radiative pressure of an electromagnetic wave is
$P_{\mathrm{rad}}=\frac{I_{\mathrm{f}}}{c}$, where $I_{\mathrm{f}}$
is the energy flux. Therefore, given a radiation field at the inner
and outer surfaces of the modeled inflow, we can estimate the radial
force on the inflow due to the radiative pressure.
$\frac{F_{\mathrm{rad}}}{s_{\mathrm{inflow}}}=P_{\mathrm{rad,incid}}+P_{\mathrm{rad,reflc}}-P_{\mathrm{rad,trans}}-P_{\mathrm{rad,diff}}$,
where $P_{\mathrm{rad,incid}}$, $P_{\mathrm{rad,reflc}}$,
$P_{\mathrm{rad,trans}}$, $P_{\mathrm{rad,diff}}$ are pressures of
incident, reflected, transmitted, and outwards diffuse emitting
radiation, and $s_{\mathrm{inflow}}$ is the irradiated area of
inflow. For the best inflow model, the outward radiative force
amounts to only $\sim 1.2-1.9\%$ of the gravitational force, where
the uncertainty mostly comes from different assumptions of the
angular distribution of reflected and diffuse emitting radiation.

If the gravitational force and the radiative resistance are the only
forces exerted on the inflow, the motion will be approximately
free-fall. According to the current estimated distance and kinetic
energy of the inflow, we find that the radial velocity equals zero
at a distance of about 1.5~pc or $\approx 1,500~R_{\mathrm{g}}$.
This is similar to the dust sublimation radius $R_{\mathrm{sub}} \approx
2,000~R_{\mathrm{g}}$ estimated using the observed
luminosity\cite{Barvainis1987} of J1035+1422, suggesting the inner
surface of the dusty torus as a natural embarkation point of the
inflow.

\subsection{Mass flux rate of the inflow}

The simplest picture of the inflow detected in redshifted BALs is
discrete clouds. In this picture, the mass inflow rate can be
estimated as $\dot{M}_{\mathrm{inflow}}= \mu
m_{\mathrm{p}}N_{\mathrm{H,inflow}}f_{\mathrm{f}}4 \pi
d_{\mathrm{inflow}}^2 \Omega_{\mathrm{inflow}}/t_{\mathrm{in}}$,
where $\mu\approx 1.4$ is the mean atomic mass per proton,
$m_{\mathrm{p}}$ is the mass of a proton, $f_{\mathrm{f}}$ is the
local filling factor, $\Omega_{\mathrm{inflow}}$ is the unknown
global covering factor of the inflow structure, and
$t_{\mathrm{in}}$ is the in-falling time scale, i.e.
$d_{\mathrm{inflow}}/v_{\mathrm{in}}$. Using the optical depth
$\tau(v)$-weighted mean of $C_{\mathrm{f}}(v)$ and $v$ in
Eq.\ref{eq1} as $f_{\mathrm{f}}$ and $v_{\mathrm{in}}$,
respectively, we have $\dot{M}_{\mathrm{inflow}}\approx
61~\Omega_{\mathrm{inflow}}~M_{\mathrm{\odot}}~\mathrm{yr}^{-1}$,
where $\Omega_{\mathrm{inflow}}$ is roughly equal to
$\Omega_{\mathrm{torus}}$ at $\sim 0.6$ under the assumption that
the inflow originates from the inner surface of the dusty torus.

An alternative picture is that the inflow structure is a continuous
layer that our LOS intercepts with an angel of $i_{\mathrm{in}}$.
The observed red-shift velocity is
$v_{\mathrm{in,obs}}=v_{\mathrm{in}}\cos i_{\mathrm{in}}$.
The mass inflow rate is estimated by calculating the amount of gas passing trough a given cross section at $d_{\mathrm{inflow}}$ in the layer in unit time:\\
\parbox{10cm}{\begin{eqnarray*} \dot{M}_{\mathrm{inflow}} &=& \mu m_{\mathrm{p}} 2\pi d_{\mathrm{inflow}}\int_{h} n_{\mathrm{H}} f_{\mathrm{f}}(h) v_{\mathrm{in}}(h)\ dh \\ &=& \mu m_{\mathrm{p}} 2\pi d_{\mathrm{inflow}}\int_{l} n_{\mathrm{H}} f_{\mathrm{f}}(l) v_{\mathrm{in,obs}}/\cos i_{\mathrm{in}} dl\sin i_{\mathrm{in}} \\ &=& \mu m_{\mathrm{p}} 2\pi d_{\mathrm{inflow}}\int^{v_1}_{v_0} dN_{\mathrm{H}}/dv_{\mathrm{in,obs}} C_{\mathrm{f}}(v_{\mathrm{in,obs}}) v_{\mathrm{in,obs}} dv_{\mathrm{in,obs}}\tan i_{\mathrm{in}} \end{eqnarray*}} \hfill
\parbox{1cm}{\begin{eqnarray}\label{eq2}\end{eqnarray}},\\
where $h$ is the height of the inflow layer at a distance
$d_{\mathrm{inflow}}$, $l$ is the length along the LOS, and $v_0$
and $v_1$ are the observed minimum and maximum radial velocities in
the redshifted trough, respectively. The result is
$\dot{M}_{\mathrm{inflow}}\approx 30~\tan
i_{\mathrm{in}}~M_{\mathrm{\odot}}~\mathrm{yr}^{-1}$. However, since
only the medium in the LoS is detected, the entire structure of the
inflow and its configuration relative to the accretion disk and
torus remains to be explored.

\section{Blueshifted BAL Outflow as a Shielding Medium}\label{outflow}

\subsection{SED-constrained blueshifted BAL outflow model}

Using the post-C$^{3+}$ region in primordial photo-ionization models
to explain the physical conditions of inflow in J1035+1422, we
suggest that the radiation illuminating the inflow is very different
from the original radiation from the central engine. In Extended
Data Fig. \ref{Fig9_simu} panel (c), we show the transmitted SED
passing through the C$^{3+}$ region for the primordial model with
$\log U=0.5$ and $\log n_{\mathrm{H}}(\mathrm{cm}^{-3})=7$. The most
prominent difference is the absence of photons with $h\nu >
48~\mathrm{eV}$ ($\lambda< 260~\mathrm{\AA}$). This SED may be a
good approximation to the actual incident radiation illuminating the
inner surface of the inflow.

The difference between the typical quasar SED and the SED required
by the inflow model is so significant that we have to postulate that
there is some kind of shielding gas obscuring the central engine, as
seen from the inflow (this is preferred over an alternative
assumption that the central engine of our object is intrinsically
abnormal). In the aspect of absorbing high energy photons, the
shielding gas is equivalent to the C$^{3+}$ and pre-C$^{3+}$ regions
(the region in front of C$^{3+}$ region where even higher-ionization
ions dominate) in the primordial models.

The blueshifted BALs consisting of C IV, Al III, Mg II doublets and
He I* $\lambda 10830$ tracing a massive outflow presents a good
candidate for the shielding gas. The absence of H I Balmer
absorptions in the blueshifted BAL system implies that the outflow
is more highly ionized than the BAL inflow, and thus is probably
closer to the central engine than the latter. In addition, the
residual flux under the blueshifted C IV BAL trough approach zero at
the deepest point, indicating that the absorber fully cover at least
the inner part of the continuum source along the LOS. Therefore, as
seen from the inflow, the object appears to be a blueshifted LoBAL
quasar.

BAL quasars are known to be weak in the soft X-ray
band\cite{Brandt2000} due to strong absorption\cite{Green2001}. The
X-ray flux of our inflow-illuminating radiation is also expected to
be heavily depressed. The quantity $\Delta \alpha_{\mathrm{OX}}=
\alpha_{\mathrm{OX}}({\rm observed})-\alpha_{\mathrm{OX}}$ was often used
to characterize the X-ray weakness of the BAL quasars in their
sample compared with non-BAL quasars, where
$\alpha_{\mathrm{OX}}({\rm observed})$ is the logarithm of the ratio
between the observed monochromatic luminosities $L_{\nu}$ at 2 keV
and $2500~\mathrm{\AA}$, and $\alpha_{\mathrm{OX}}$ is the same
quantity for a typical non-BAL quasar\cite{Gibson2009}. The X-ray
depression is found to be more severe in LoBAL quasars than in HiBAL
quasars, because the sub-sample of LoBAL quasars has systematically
smaller $\Delta \alpha_{\mathrm{OX}}$.

The predicted flux depression at 2 keV is plotted in Extended Data
Fig. \ref{Fig9_simu} panel (c), where the average values of $\Delta
\alpha_{\mathrm{OX}}$ for HiBAL and LoBAL samples has been assumed,
respectively. The anticipated inflow-illuminating flux is even lower
than the average value for LoBAL sample. However, due to the fact
that the majority of BAL quasars remain undetected in X-ray as yet,
these `average' values are likely to be some kind of upper limits.

Further quantitative analysis suggests that, for the post-C$^{3+}$
inflow model, the transmitted SED of any gaseous plate with the same
$U$ and $N_{\mathrm{H}}$ as the pre-C$^{3+}$ and C$^{3+}$ regions
associated with the post-C$^{3+}$ inflow can match the incident SED
required by the inflow. For example, in Extended Data Fig.
\ref{Fig9_simu} panel (b) we present the shielding gas model with
$\log U=0.5$, $\log n_{\mathrm{H}}(\mathrm{cm}^{-3})=9.5$, and $\log
N_{\mathrm{H}}(\mathrm{cm}^{-2})=23.5$, through which the
transmitted SED matches the incident SED required by the inflow
model with $\log n_{\mathrm{H}}(\mathrm{cm}^{-3})=7$ shown in
Extended Data Fig. \ref{Fig9_simu} panel (a), noticing that for the
associated pre-C$^{3+}$ and C$^{3+}$ regions $\log U=0.5$ and $\log
N_{\mathrm{H}}(\mathrm{cm}^{-2})=23.56$. Since $U$ is fixed, the
denser the shielding gas is, the cLOSer it is to the central engine.
In the given example, as $n_{\mathrm{H}}$ for the shielding gas
model is 2.5 orders of magnitude larger than the inflow model, the
corresponding outflow is inferred to be much closer to the central
engine, at $\sim 82~R_{\mathrm{g}}$.

\subsection{Comparison with the observation of blueshifted BAL system}

However, these models seem inconsistent with the measurements for
the blueshifted BAL systems. Since the blueshifted C IV BAL trough
is saturated, we can only derive rather loose constraints that
$N_{\mathrm{col}}(\mathrm{C}^{3+}_{\mathrm{ground}})$ should not be
smaller than a few $10^{15}~\mathrm{cm}^{-2}$. If we assume that the
outflow fully obscures the continuum source and the emission region
as the C IV trough indicates, we find that the blueshifted He I*
$\lambda 10830$ are unsaturated, making the measurements of $\tau
(v)(\mathrm{He I*} \lambda 10830)$ feasible. Therefore, we have
$N_{\mathrm{col}}(\mathrm{He}^0 2^3\mathrm{S})=6.46\pm 1.48\times
10^{13}~\mathrm{cm}^{-2}$. However, the SED-constrained outflow
model predicts a much larger $N_{\mathrm{col}}(\mathrm{He}^0
2^3\mathrm{S})$. For the outflow model plotted in Extended Data Fig.
\ref{Fig9_simu} panel (b), $N_{\mathrm{col}}(\mathrm{He}^0
2^3\mathrm{S})>10^{15}~\mathrm{cm}^{-2}$. Alternatively, if we use
the measured $N_{\mathrm{col}}(\mathrm{He}^0 2^3\mathrm{S})$ to
define the thickness of the outflow ($N_{\mathrm{H}}(\mathrm{He
I*})$ in Extended Data Fig. \ref{Fig9_simu} panel (b)), the
shielding gas would be overly thin. The transmitted SED at
$N_{\mathrm{H}}(\mathrm{He I*})$ is plotted in Extended Data Fig.
\ref{Fig9_simu} panel (c), in which the amount of residual high
energy photons are still large enough to generate C$^{3+}$ far more
than that measured in the redshifted system.

A possible explanation for such an inconsistency is the geometry
issue. Since the outflow in our objects is considered cLOSe to the
very center of the quasar nucleus, where the nature (e.g. the
thermal structure) of the inner accretion disk or the origin of the
X-ray is not fully understood, and the assumption that the covering
factor of the blueshifted BAL system is wavelength-independent may
be questionable\cite{Shi2017}. If $C_{\mathrm{f}}$ decreases as
wavelength increases, our current measurements for
$N_{\mathrm{col}}(\mathrm{He}^0 2^3\mathrm{S})$ can be a significant
underestimation. The true value of $N_{\mathrm{col}}(\mathrm{He}^0
2^3\mathrm{S})$ may be consistent with the SED-constrained outflow
thickness.

Besides the geometry issue, the metallicity, which has been found to
be super-solar and varying considerably in outflows\cite{Wang2012} may also be an explanation for the inconsistency. While the
metal abundance has little effect on the ionization structure of
He$^0(2^3\mathrm{S})$, the C$^{3+}$ region moves forward (toward the
illuminated surface) when the metal abundance increases. At solar
abundances, $N_{\mathrm{H,outflow}}$ that is determined using the
full development of C$^{3+}$ region is about 3 times as large as
that determined using $N_{\mathrm{col}}(\mathrm{He}^0
2^3\mathrm{S})$, while for $Z=10~Z_{\odot}$ these two values are
about the same. In Extended Data Fig. \ref{Fig10_metallicity}, we
plot the transmitted SED through gas of $N_{\mathrm{H}}(\mathrm{He
I^*})$ for $Z=0.1,1,10~Z_{\odot}$, respectively. Obviously, with
increasing metallicity, the transmitted SED tends to resemble the
incident SED, as expected by our inflow models.

\subsection{Mass flux rate of the blueshifted BAL outflow}

Following the discussion of a continuous layer as the inflow model, the mass flux rate of outflow can also be expressed as\\
\parbox{10cm}{\begin{eqnarray*} \dot{M}_{\mathrm{outflow}} &=& \mu m_{\mathrm{p}} 2\pi d_{\mathrm{outflow}}\int_{h'} n_{\mathrm{H}} f_{\mathrm{f}}(h') v_{\mathrm{out}}(h')\ dh' \\ &=& \mu m_{\mathrm{p}} 2\pi d_{\mathrm{outflow}}\int_{l'} n_{\mathrm{H}} f_{\mathrm{f}}(l') v_{\mathrm{out,obs}}/\cos i_{\mathrm{out}} dl\sin i_{\mathrm{out}} \\ &=& \mu m_{\mathrm{p}} 2\pi d_{\mathrm{outflow}}\int^{v_1'}_{v_0'} dN_{\mathrm{H}}/dv_{\mathrm{out,obs}} C_{\mathrm{f}}(v_{\mathrm{out,obs}}) v_{\mathrm{out,obs}} dv_{\mathrm{out,obs}}\tan i_{\mathrm{out}} \end{eqnarray*}} \hfill
\parbox{1cm}{\begin{eqnarray}\label{eq3}\end{eqnarray}},\\
where $h'$ is the height of the outflow layer at the distance
$d_{\mathrm{outflow}}$, $l'$ is the length along the LOS,
$i_{mathrm{out}}$ is the angle between the true outflow velocity and
our LOS, and $v_0'$ and $v_1'$ are the observed minimum and maximum
radial velocities of the blueshifted C IV trough, respectively. As
$C_{\mathrm{f}}(v_{\mathrm{out,obs}})=1$ for the blueshifted C IV
BAL, we have $\dot{M}_{\mathrm{outflow}}\approx
0.05~d_{\mathrm{outflow}}(R_{\mathrm{g}})~\tan
i_{\mathrm{out}}~M_{\mathrm{\odot}}~\mathrm{yr}^{-1}$ for the best
inflow model, assuming the outflow structure to be axisymmetric.
Though $d_{\mathrm{outflow}}$ can hardly be constrained by the
observation, the value of $d_{\mathrm{outflow}}$ in units of
$R_{\mathrm{g}}$ ($d_{\mathrm{outflow}}(R_{\mathrm{g}})$) is
believed to be between $\sim 100$ and $\sim 1,000$.

\section{BAL Inflow and Outflow in Emission Lines}

Only when intercepting our LOS will the inflow or outflow gas be
observed as redshifted or blueshifted BAL systems. However, the
emission from the inflows and outflows should be detectable in any
direction. Our simulation can also predict the surface emissivity
(radiative energy output from unit area) for various emission lines
in both of the outflows and inflows. In the outflow's emission
lines, we find little variation in the emissivity within the range
of distance $100 - 1,000~\mathrm{R_{\mathrm{g}}}$ with [O III] as
the only exception, which is strongly dependent on the flows'
density. The ratios between the emissivity in the inflow and that in
the outflow are found to depend on the ionization state, as we
expect, according to the ionization structure. For C IV the
emissivity in an inflow is 3 orders of magnitude less than that in
an outflow, while for Mg II the emissivity in the inflow is about 5
times larger than that in the outflow. In the case of H$\alpha$, the
emissivity in the inflow is comparable to that in the outflow.
Detection of the expected quasar inflows in low-ionization emission
lines will be presented in a forthcoming paper.

In Supplementary Table 1, we present the equivalent widths (EWs) of
various emission lines predicted for inflows and outflows. Assuming
global covering factors $\Omega$ of 0.6 for inflows and 0.4 for
outflows, the EWs for both outflows' and inflows' emission for
H$\alpha$ are of several tens \AA, which is comparable to the broad
emission of typical quasars and thus supposed to be detectable.
Actually, we do find the H$\alpha$ peak in J1035+1422 much wider
than that in the SDSS composite\cite{VandenBerk2001}, which could be
an emission signature from outflows and inflows (Extended Data Fig.
\ref{Fig11_PreHalpha}). The excess parts on the blue and red may be
originated from the accretion disk as in a small fraction ($\sim
4\%$) of AGNs that show a `double-peaked' profile in low-ionization
broad emission lines\cite{Eracleous2004}. Detailed discussion of
emission lines from inflows and outflows will be presented in a
companion paper.

\section{Data availability. }

The observations discussed in this paper were mostly made using the
P200 under Telescope Access Programme (TAP).

\end{methods}

\clearpage
%%%%%%%%%%%%%%%%%
\section*{Additional references}

\clearpage

\renewcommand{\figurename}{\textbf{Extended Data Fig.}}
\renewcommand{\tablename}{\textbf{Extended Data Table}}

\setcounter{figure}{0}

\newpage

\begin{figure}
\begin{center}
\includegraphics[width=0.7\textwidth]{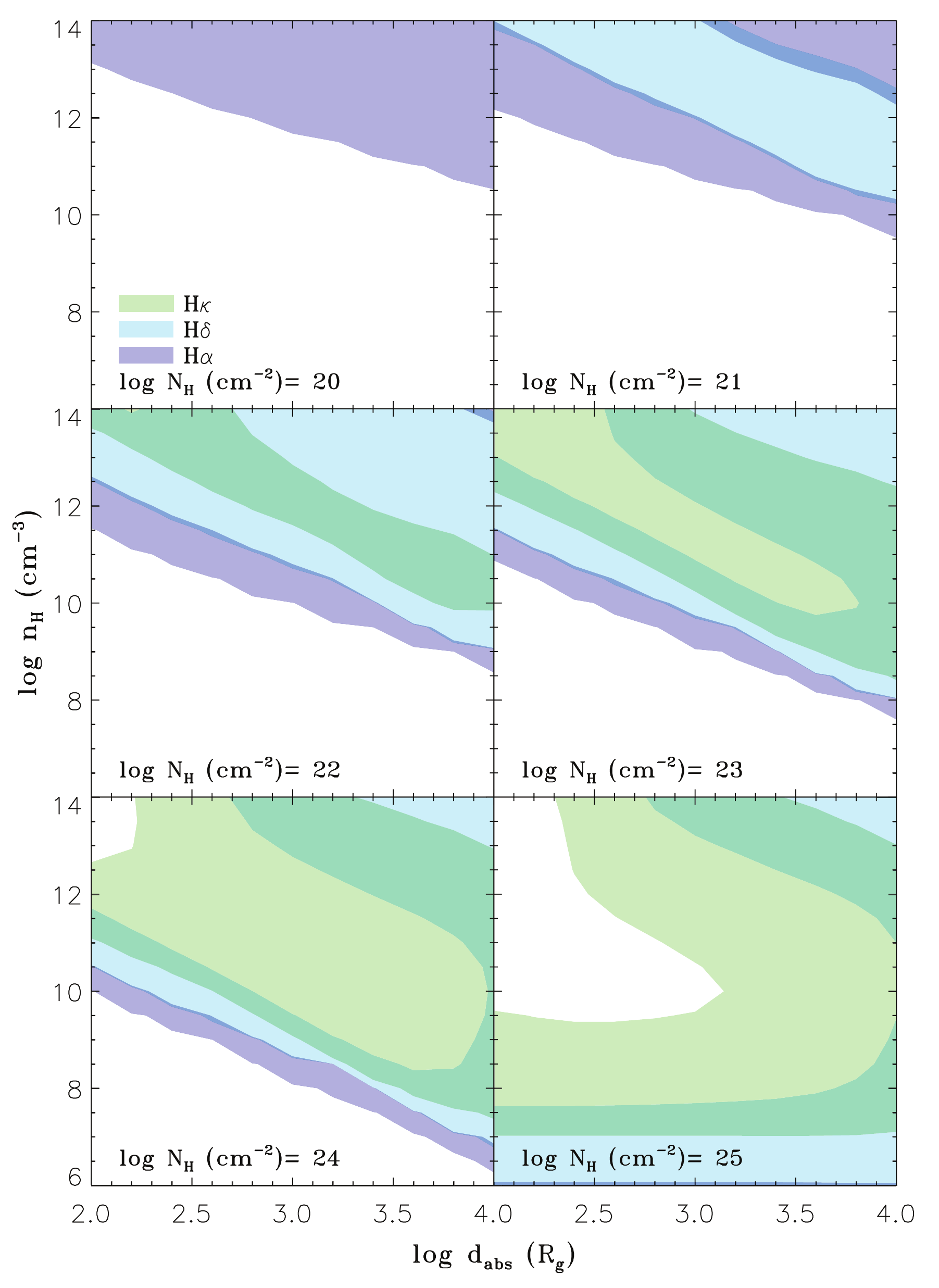}
\end{center}
\caption[]{\textbf{Sensitive range of the BALs in the hydrogen
Balmer series H$_{n+2}$ for gas in the vicinity of a SMBH.}
Assuming a black hole of $M_{\bullet}=10^9~M_{\odot}$ accreting
at an Eddington ratio of 0.1,
the column densities of H$_{n=2}$ at the $n=2$ level $N_{\mathrm{col}}(\mathrm{H}^0_{n=2})$
are evaluated by photo-ionization simulations for gas of various
density $n_{H}$, total column density $N_{H}$
and distance (in units of $R_{\rm g}$). Assuming a Gaussian
velocity dispersion (${\rm FWHM}=3,000~{\rm km~s}^{-1}$), a specified line would
be considered to be sensitive in measuring the ionic column density as long as
the optical depth is in the range of 0.05--3 at the line center. The
colored area shows the sensitive range for each of the individual
lines of only H$\alpha$, H$\delta$, and H$\kappa$ for clarity
($\lambda_{\rm rest}=6,564$, 4,102 and 3,750~\AA, respectively).}
\label{Fig1_Hydrogen}
\end{figure}

\clearpage

\begin{figure}
\begin{center}
\includegraphics[width=0.7\textwidth]{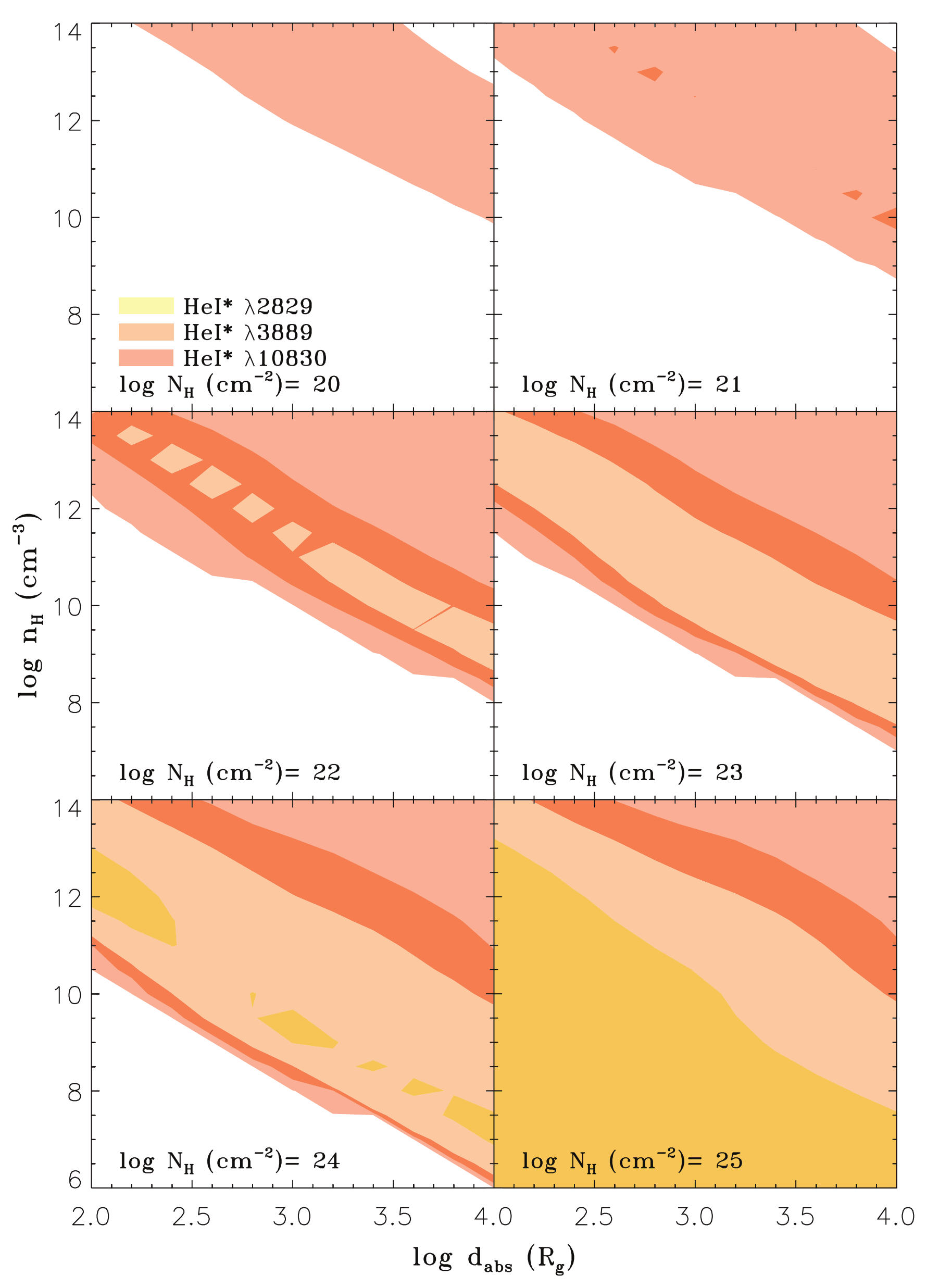}
\end{center}
\caption[]{Same as Extended Data Fig. \ref{Fig1_Hydrogen}, but for
the He I$^{*}_{n}$ BALs. } 
\label{Fig2_Helium}
\end{figure}

\clearpage

\begin{figure}
\begin{center}
\includegraphics[width=0.65\textwidth]{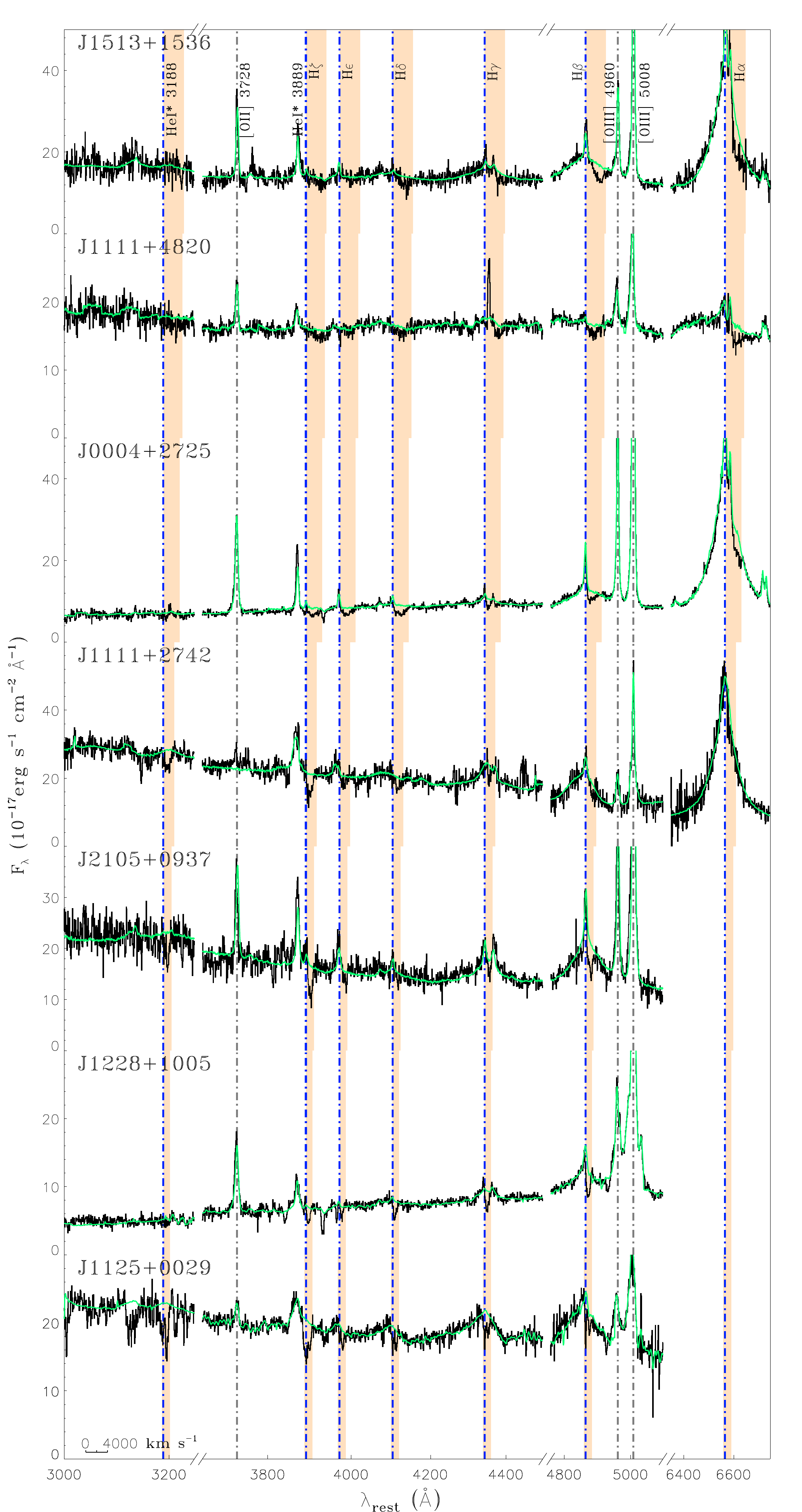}
\end{center}
\caption[]{SDSS observed spectra of the remaining seven quasars of
our sample with pure redshifted BALs (or mini-BALs) in the H Balmer
and meta-stable He I* plotted in their rest frames. The systemic
redshifts are determined from narrow emission lines including [O~II]
(gray dotted-dashed vertical lines). As in Figure\,1 of the main
text, the blue dash-dotted lines mark the rest wavelengths of the H
Balmer and He I* transitions. The wavelength ranges of the
absorption lines are shaded in orange.} \label{Fig3_sample}
\end{figure}

\clearpage

\begin{figure}
\begin{center}
\includegraphics[width=1.0\textwidth]{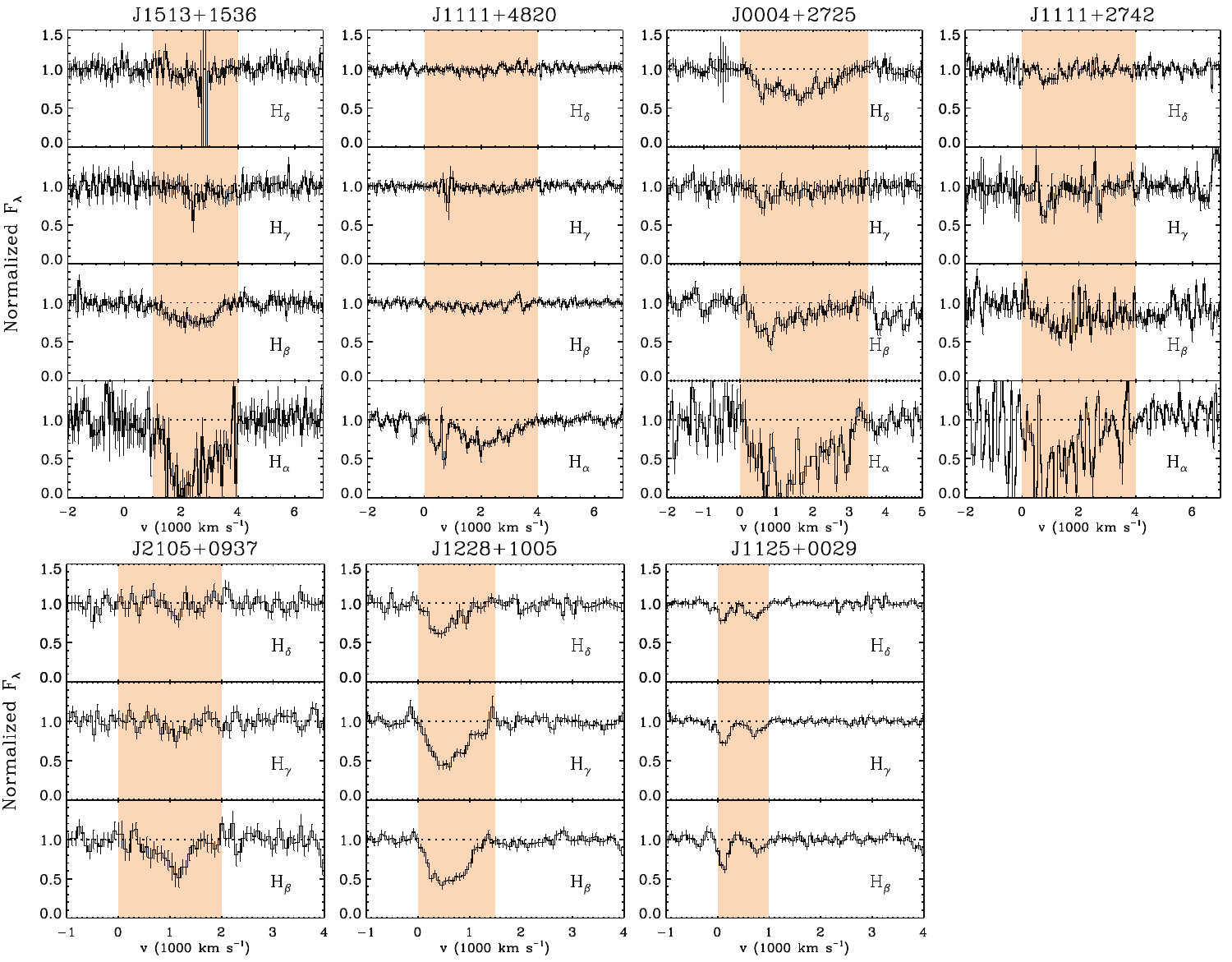}
\end{center}
\caption[]{Selected Balmer absorption lines (shaded in orange) on
the SDSS spectra of the seven quasars in Extended Data Fig.
\ref{Fig3_sample} plotted in their common velocity space. The data
are normalized by the continuum after subtracting the best-fit
emission line models (as for J1035+1422 in Figure
\ref{Figure2_BALs}). The upper panels show four bona fide BAL
quasars with absorption troughs spanning a large range of velocities
from $\sim 0$ to $\sim 4,000~{\rm km~s}^{-1}$, significantly broader than
the BAL definition criterion of $2,000~{\rm km~s}^{-1}$). The absorption
troughs of the remaining three quasars in the lower panels have
widths $\sim 1,000-2,000~{\rm km~s}^{-1}$ and are formally classified as
`mini-BALs'.  } \label{Fig4_sample}
\end{figure}

\clearpage

\begin{figure}
\begin{center}
\includegraphics[width=\textwidth]{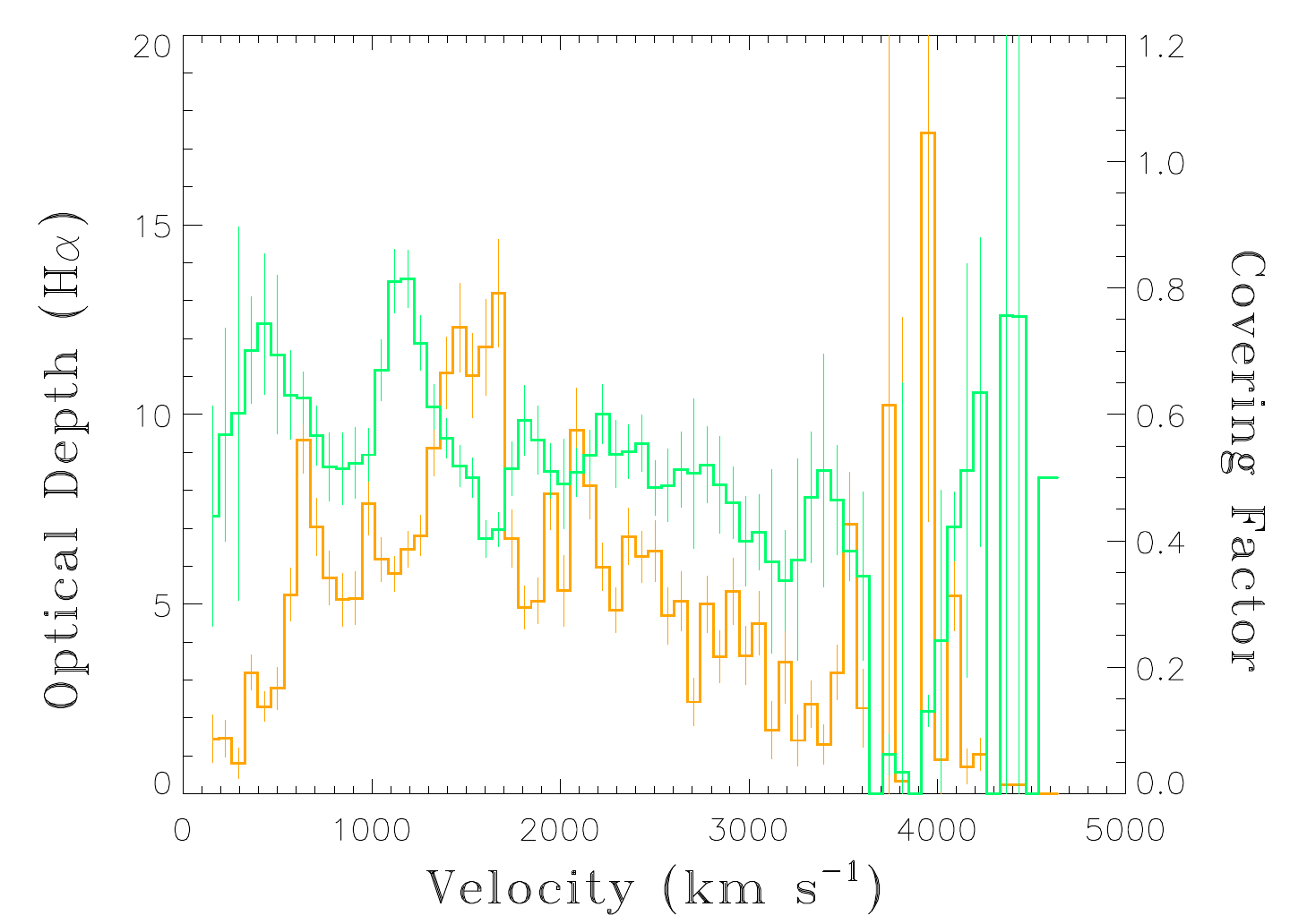}
\end{center}
\caption[]{Optical depth (orange) and the covering factor (green) of
the redshifted H$\alpha$ BAL,
derived from the continuum-normalized spectrum,
as a function of velocity shift with respect to the quasar's rest-frame.}
\label{Fig5_Halpha}
\end{figure}

\clearpage

\begin{figure}
\begin{center}
\includegraphics[width=\textwidth]{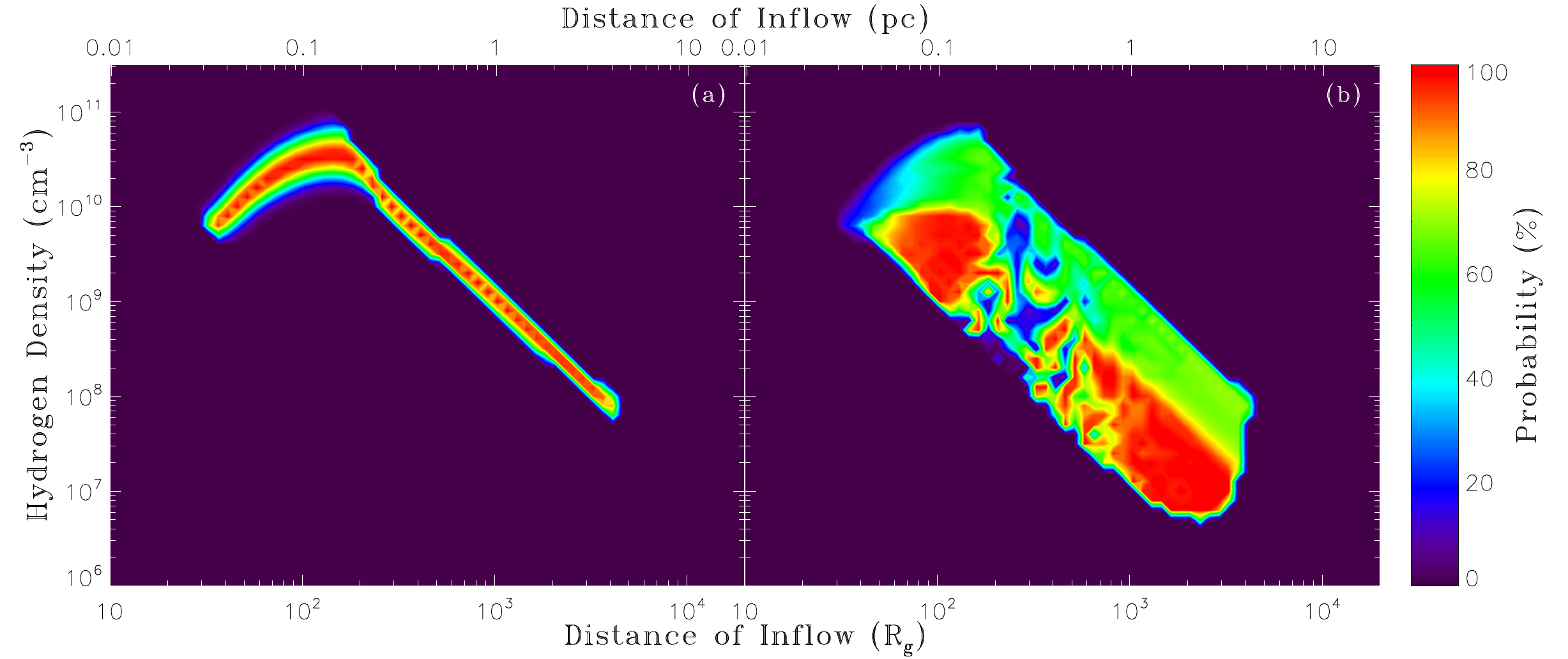}
\end{center}
\caption[]{Probability density distribution in the parameter space
of the total hydrogen density $n_{\mathrm{H}}$ and the distance from
the central engine $d_{\mathrm{inflow}}$ for the inflow models (see
also Figure \ref{Figure3_probability} of the main text). In panel
(a) the simplest primordial models are applied, and the redshifted
H\,I Balmer and He\,I* BALs are employed to evaluate the probability
density. However, the highly probable models predict much higher
column densities of $\mathrm{C}^{3+}$ ions
$N_{\mathrm{col}}(\mathrm{C}^{3+}_{\mathrm{ground}})$ than that
evaluated from the redshifted C\,IV BAL trough. To resolve this
problem, only the region beyond the C$^{3+}$ region (post C$^{3+}$
region in short) in the primordial models is used to describe the
inflow gas, and probability density is recalculated by including C
IV. The results of these refined model calculations are displayed in
panel (b), where the probability density shows two peaks around
$n_{\mathrm{H}}\approx 10^7~\mathrm{cm}^{-3}$ and
$d_{\mathrm{inflow}}\approx 2,000~R_{\mathrm{g}}$, and
$n_{\mathrm{H}}\approx 10^{9.5}~\mathrm{cm}^{-3}$ and
$d_{\mathrm{inflow}}\approx 100~R_{\mathrm{g}}$, respectively.}
\label{Fig6_prob}
\end{figure}

\clearpage

\begin{figure}
\begin{center}
\includegraphics[width=\textwidth]{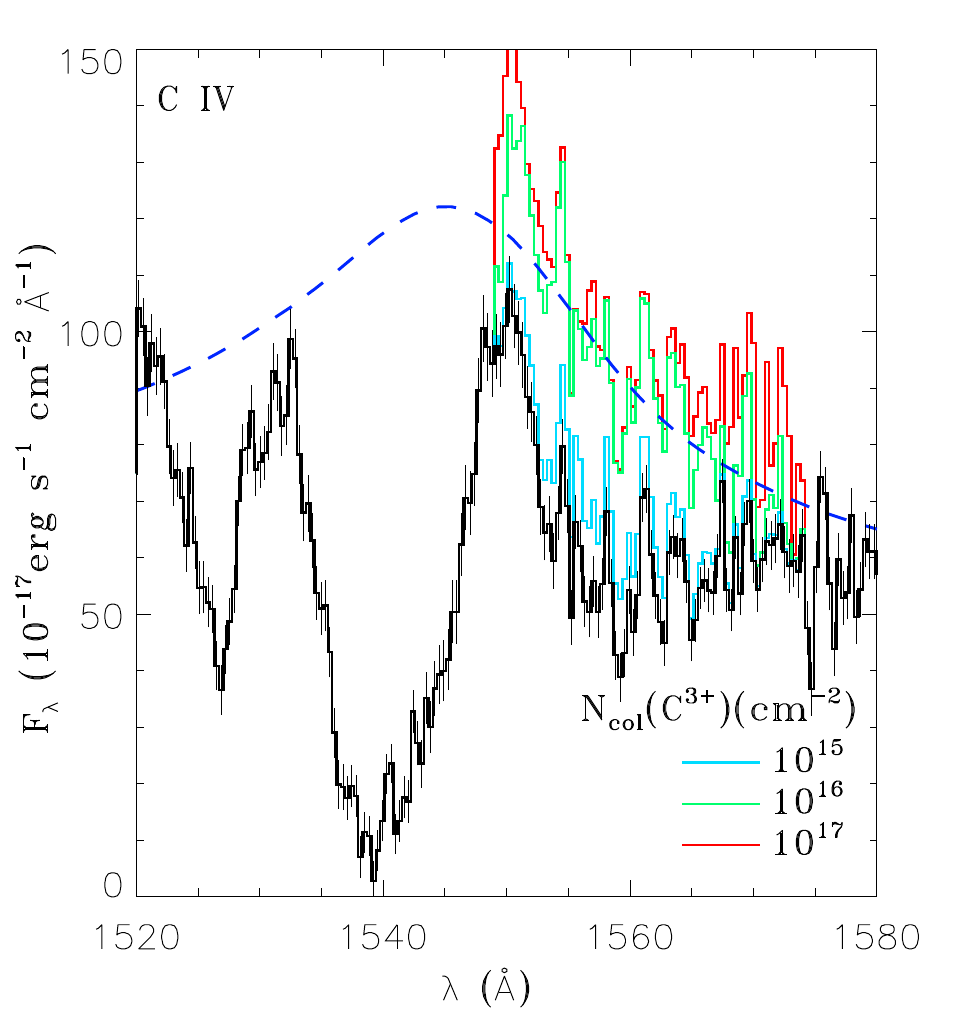}
\end{center}
\caption[]{Recovered spectra of C IV corrected for the redshifted absorption
assuming $\mathrm{C}^{3+}$ ion column densities of $\log
N_{\mathrm{col}}(\mathrm{C}^{3+}_{\mathrm{ground}})(\mathrm{cm}^{-2})=15$
(cyan), $16$ (green) and $17$ (red) in the quasar's rest-frame.
Compared to the best-fit SDSS composite spectrum (blue dashed line),
the recovered flux is much too weak for the absorption with
$\log N_{\mathrm{col}}(\mathrm{C}^{3+}_{\mathrm{ground}})(\mathrm{cm}^{-2})=15$,
while it is too high showing two extra deceptive peaks at
$\sim 1550$ and $1570~\mathrm{\AA}$
for the absorption with
$\log N_{\mathrm{col}}(\mathrm{C}^{3+}_{\mathrm{ground}})(\mathrm{cm}^{-2})=17$.
The absorption model with $\log
N_{\mathrm{col}}(\mathrm{C}^{3+}_{\mathrm{ground}})(\mathrm{cm}^{-2})\approx 16$
predicts unabsorbed flux reasonably consistent with the composite
spectrum, and is thus adopted.}
\label{Fig7_CIV}
\end{figure}

\clearpage

\begin{figure}
\begin{center}
\includegraphics[width=\textwidth]{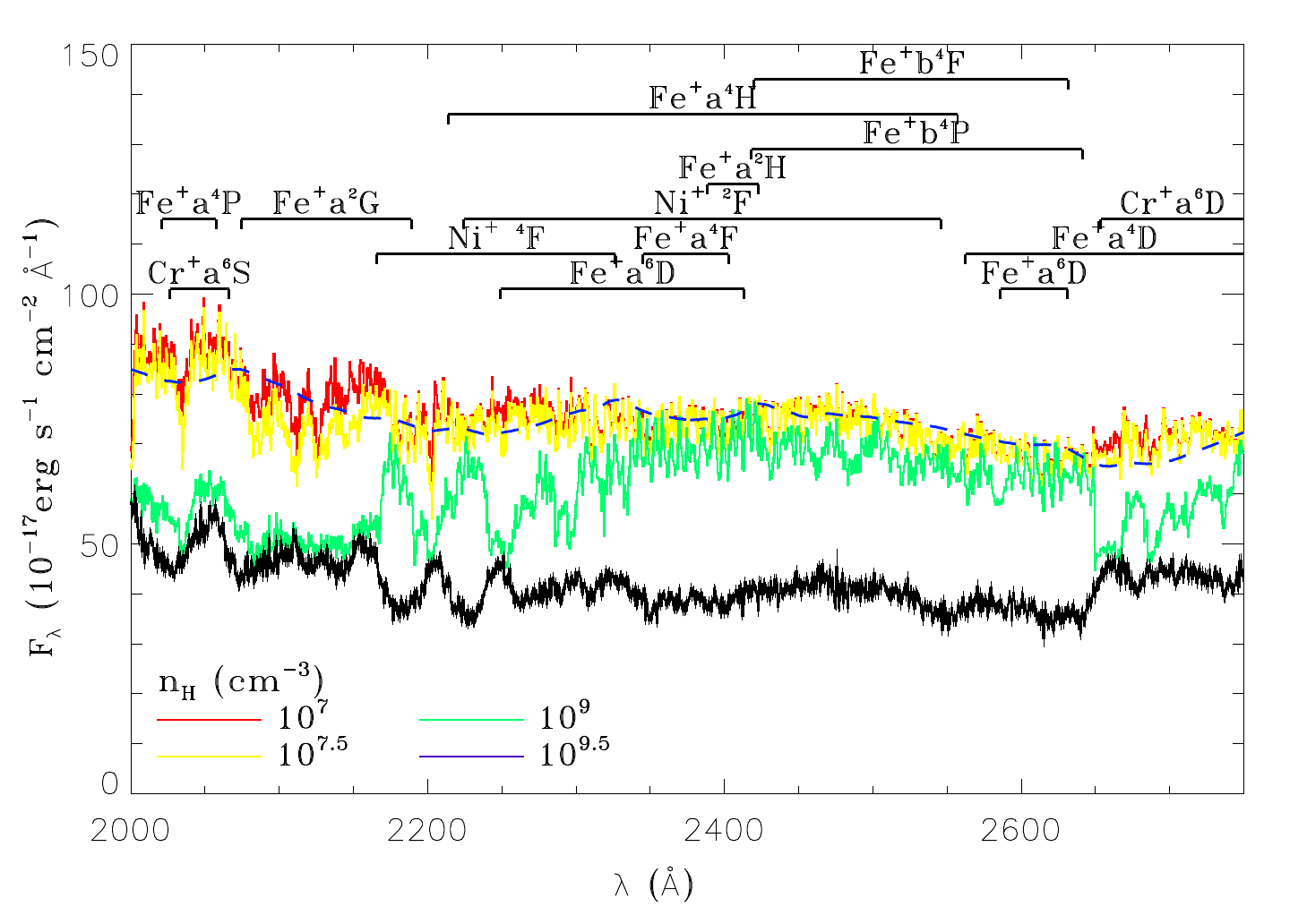}
\end{center}
\caption[]{Absorption-corrected UV Fe II spectra between 2000
and $2750~\mathrm{\AA}$ for the post-C$^{3+}$ inflow models with
$\log n_{\mathrm{H}}(\mathrm{cm}^{-3})=$7 (red), 7.5 (yellow), 9
(green), and 9.5 (violet) in the high `probability' zone of Extended
Data Fig. \ref{Fig6_prob} panel (b). Compared with the best-fit
composite (blue dashed), the models with higher densities can be
clearly ruled out.} \label{Fig8_UVFeII}
\end{figure}

\clearpage

\begin{figure}
\begin{center}
\includegraphics[width=0.4\textwidth]{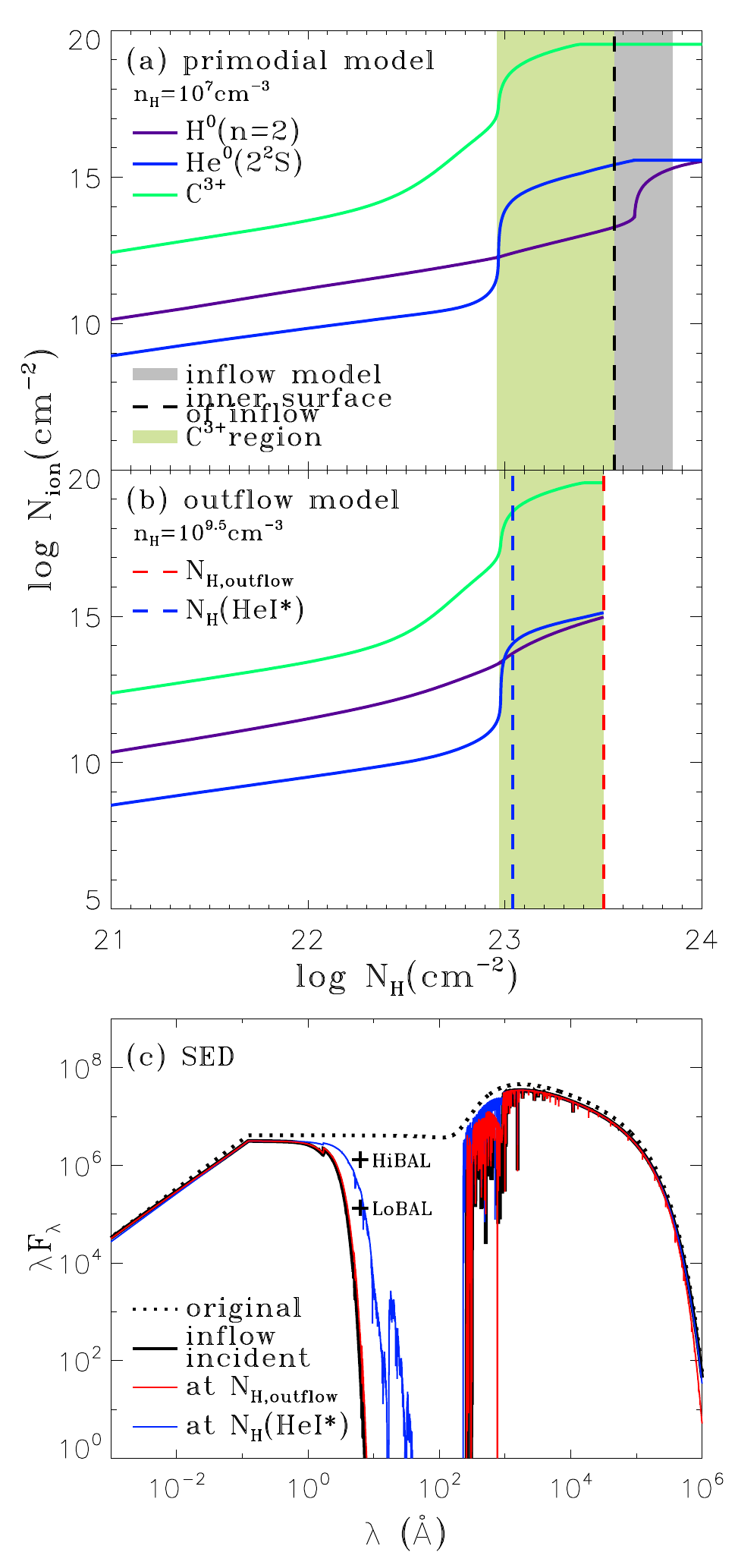}
\end{center}
\caption[]{Photo-ionization model for the inflow and outflow. Panel
(a) shows the ionization structure of a primordial model with
$n_{\mathrm{H}}=10^7~\mathrm{cm}^{-3}$ and $U=10^{0.5}$, which is
directly illuminated by the central continuum source of the quasar.
If integrated from the illuminated surface, the model with
$N_{\mathrm{col}}(\mathrm{H}^0_{n=2})$ and
$N_{\mathrm{col}}(\mathrm{He}^0 2^3\mathrm{S})$ comparable to the
measurements predicts
$N_{\mathrm{col}}(\mathrm{C}^{3+}_{\mathrm{ground}})>10^{19}~\mathrm{cm}^{-2}$,
far from the estimated
$N_{\mathrm{col}}(\mathrm{C}^{3+}_{\mathrm{ground}})$ in the
redshifted BAL. An alternative solution is that the inflow in fact
corresponds to the gas behind the C$^{3+}$ region (the light green
area where C$^{3+}$ and other high-ionization ions dominate), which
is shown as the grey area. In such a picture, the outflow is
suggested to play an equivalent role as the C$^{3+}$ region in panel
(a) in eliminating high energy ionizing photons. In panel (b), we
plot the ionization structure for the outflow with
$n_{\mathrm{H}}=10^{9.5}~\mathrm{cm}^{-3}$ and $U=10^{0.5}$. The
requirement for the transmitted radiation (which should have the
same SED as the incident radiation on inflow) could constrain the
thickness of outflow model. The outer surface of this model (red
dashed line) highly coincides the extension of the C$^{3+}$ region.
However, $N_{\mathrm{col}}(\mathrm{He}^0 2^3\mathrm{S})$ measured
using the blueshifted He I* $\lambda 10830$ defines a thinner
outflow gas (see the blue dashed line) if we assume the covering
factor $C_{\mathrm{f}}$ is wavelength-independent. In panel (c), we
plot the transmitted SEDs through the SED-constrained outflow and
the $N_{\mathrm{col}}(\mathrm{He}^0 2^3\mathrm{S})$-defined outflow.
The former (red) naturally consists with the incident SED for the
inflow model, while the latter (blue) shows considerable excess in
soft X-ray which would result in a much larger
$N_{\mathrm{col}}(\mathrm{C}^{3+}_{\mathrm{ground}})$ in the inflow
than the measurement.} \label{Fig9_simu}
\end{figure}

\clearpage

\begin{figure}
\begin{center}
\includegraphics[width=\textwidth]{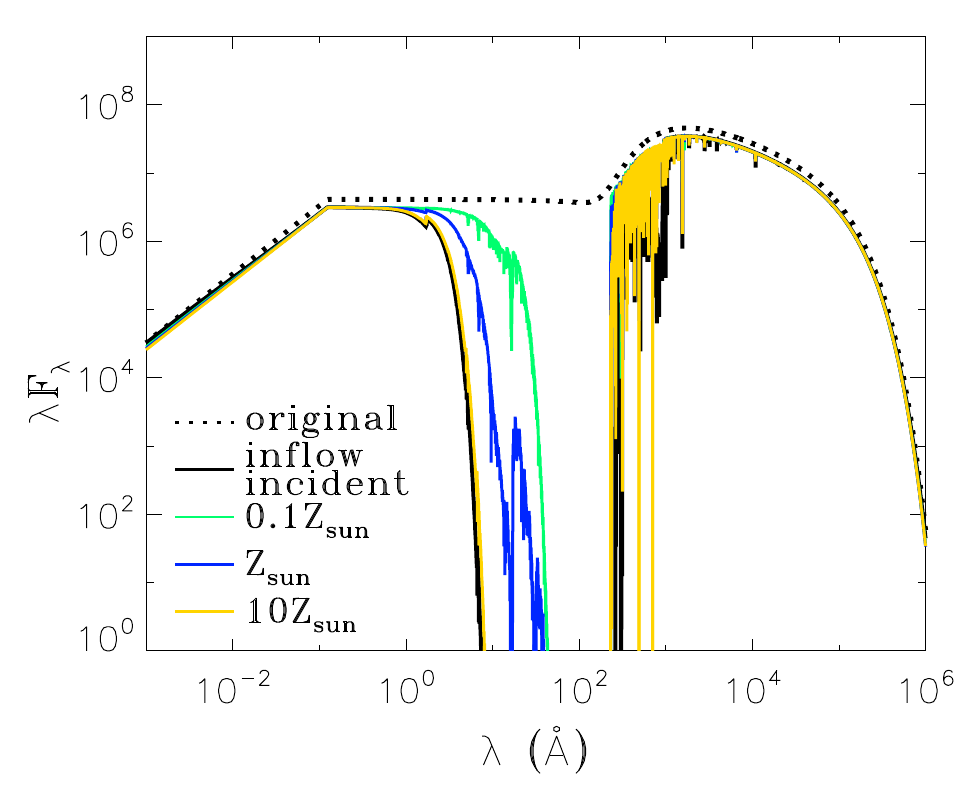}
\end{center}
\caption[]{Transmitted SEDs through a
$N_{\mathrm{col}}(\mathrm{He}^0 2^3\mathrm{S})$-defined outflow
model with $n_{\mathrm{H}}=10^{9.5}~\mathrm{cm}^{-3}$, $U=10^{0.5}$
and different metallicities. The SED sensitively depends on the
metallicity.
As the metallicity increases to from solar to
$10~Z_{\mathrm{\odot}}$, the transmitted SED seems to match
the incident SED required by the inflow model, suggesting that
a metal-rich outflow model could explain the measurement in both
the redshifted and blueshifted BAL systems.} \label{Fig10_metallicity}
\end{figure}

\clearpage

\begin{figure}
\begin{center}
\includegraphics[width=0.7\textwidth]{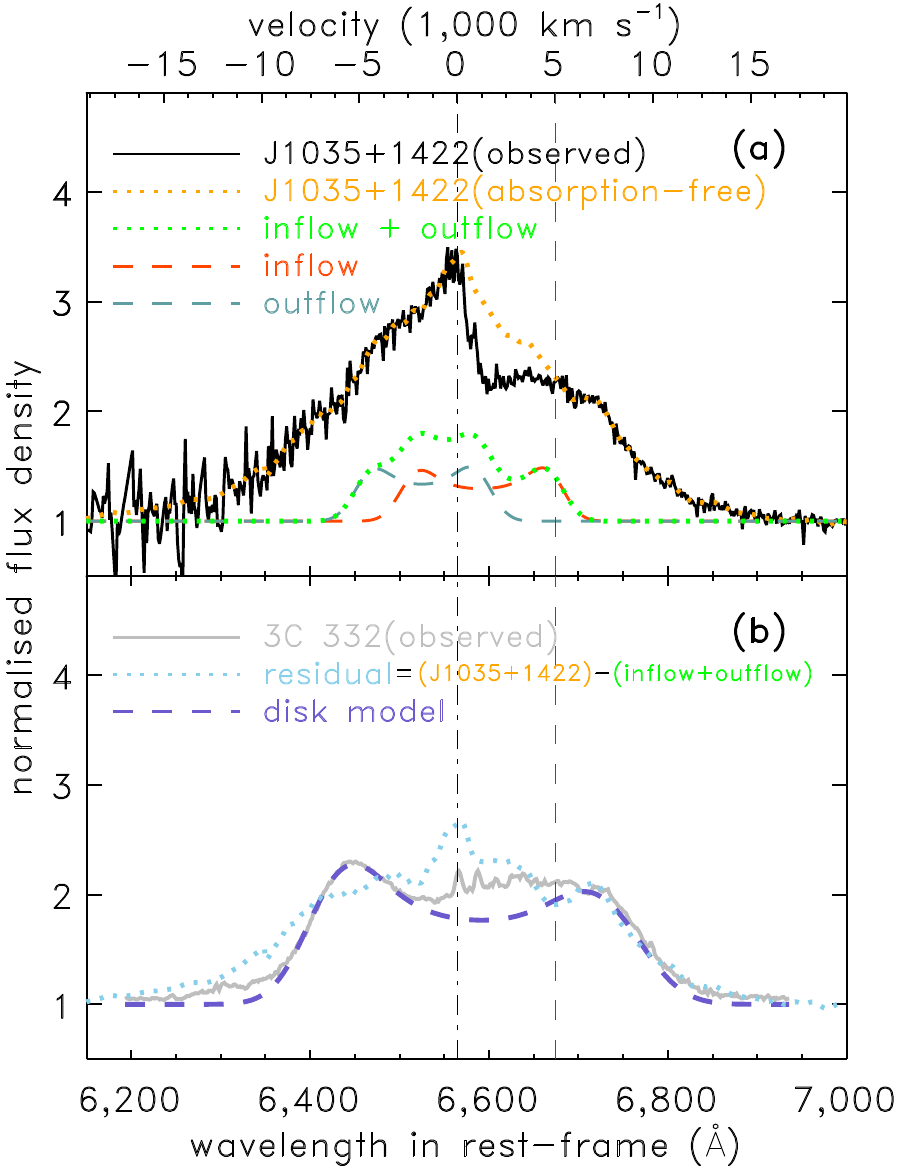}
\end{center}
\caption[]{Upper panel:
Observed (dark gray solid line) and the
absorption-corrected (orange dotted line)
broad H$\alpha$ emission line spectra of J1035+1422
normalized to the continuum.
Over-plotted for comparison are the
predicted H$\alpha$ emission lines by
the photo-ionization models from the inflow (red dashed line),
the outflow (blue dashed line),
and the total as a combination of the two
(green dotted line)
(assuming a covering factor of 0.5
($\Omega_{\mathrm{i}}=\Omega_{\mathrm{o}}=0.5$ in Supplementary
Table \ref{emissionEW}) and a radial velocity
$5,000~{\rm km~s}^{-1}$ for both the inflow and outflow.)
Clearly, the observed H$\alpha$
is significantly stronger than the model prediction.
This may be due to the over-simplicity of the models,
in which a much broader velocity range is missing.
Alternatively and more likely, the excess H$\alpha$ flux may
be contributed from the accretion disk.
The lower panel shows the residual line profile
(cyan dotted line; the zigzag
shape is caused by the oversimplified model assumption of a single
velocity instead of a large velocity gradient in reality),
which largely resembles the H$\alpha$ line observed in the well-studied disk
emitting quasar 3C\,332\cite{Eracleous2004} (gray solid line).
Note that 3C\,332 shows a significant
excess component with respect to the best-fit disk line model
(violet dashed line),
which is redshifted with a velocity range $\sim 0-5,000~{\rm km~s}^{-1}$.
This is reminiscent of the redshifted H$\alpha$ BAL
found in J1035+1422 here, suggestive of an interesting possibility
that this excess H$\alpha$ emission might originate from inflows
in 3C\,332. }
\label{Fig11_PreHalpha}
\end{figure}

\begin{table}
\centering
%\caption
{Supplementary Table 1. Predicted emission lines from the outflow
and inflow of J1035+1422 in comparison with the averaged BEL EWs of
quasars.}
{\footnotesize \label{emissionEW} \\

\begin{tabular}{lrrrr}
\hline
line                              & outflow$^{\dag}$                 &  inflow$^{\dag}$            & averaged BEL  & reference\\
                                  & \AA                              &  \AA                        &   \AA         & \cite{VandenBerk2001}\\
C IV $\lambda\lambda 1548,1550$   & $65-69~\Omega_{\mathrm{o}}$      & $0.051~\Omega_{\mathrm{i}}$ &   $23.78$     & \cite{VandenBerk2001}\\
Al III $\lambda\lambda 1855,1863$ & $2.3~\Omega_{\mathrm{o}}$        & $0.81~\Omega_{\mathrm{i}}$  &   $0.40$      & \cite{VandenBerk2001}\\
Mg II $\lambda\lambda 2796,2803$  & $5.0~\Omega_{\mathrm{o}}$        & $32~\Omega_{\mathrm{i}}$    &   $32.28$     & \cite{VandenBerk2001}\\
H$\alpha$                         & $140~\Omega_{\mathrm{o}}$        & $150~\Omega_{\mathrm{i}}$   &   $194.52$    & \cite{VandenBerk2001}\\
He I$^*\lambda 10830$             & $53-61~\Omega_{\mathrm{o}}$      & $15~\Omega_{\mathrm{i}}$    &   $36$        & \cite{Glikman2006}   \\

\hline \multicolumn{5}{}{ $^{\dag}$ Equivalent width in the quasar's
rest-frame. $\Omega_{\mathrm{o}}$ and $\Omega_{\mathrm{i}}$ are the
global covering factors of outflow and inflow, respectively.}
\end{tabular}}
\end{table}

\end{document}